\documentclass[aps,pra,reprint,superscriptaddress,nopacs,nofootinbib,longbibliography]{revtex4-1}

\usepackage{amsmath,amssymb,latexsym}
\usepackage{graphicx}
\graphicspath{{../fig/}}
\usepackage{hyperref}
\usepackage{mathrsfs}
\usepackage{amscd}
\usepackage{here}
\usepackage{color}
\usepackage{comment}
\DeclareSymbolFont{symbols}{OMS}{cmsy}{m}{n}
\DeclareSymbolFont{largesymbols}{OMX}{cmex}{m}{n}
\renewcommand{\bm}[1]{\boldsymbol #1}

\begin{document}

\title{
Higgs-mode resonance in third harmonic generation in NbN superconductors:
Multiband electron-phonon coupling, impurity scattering, and polarization-angle dependence
}

\author{Naoto Tsuji}
\affiliation{RIKEN Center for Emergent Matter Science (CEMS), Wako 351-0198, Japan}
\author{Yusuke Nomura}
\affiliation{RIKEN Center for Emergent Matter Science (CEMS), Wako 351-0198, Japan}

\begin{abstract}
We theoretically investigate the resonance of third harmonic generation (THG)
that has been observed at frequency being half of the superconducting gap in a multiband disordered superconductor NbN.
The central question is whether the dominant contribution to the THG resonance comes from the Higgs mode 
(the collective amplitude mode of the superconducting order parameter) or quasiparticle excitations. 
To resolve this issue, 
we analyze a realistic three-band model with effective intraband and interband phonon-mediated interactions together with nonmagnetic impurity scatterings. 
Using the first principles estimate of the ratio between the intraband and interband pairing interactions with multiband impurity scattering rates being varied from clean to dirty regimes,
we calculate the THG susceptibility for NbN in a channel-resolved manner
by means of the BCS and self-consistent Born approximations.
In the dirty regime, which is close to the experimental situation, 
the leading contribution is given by the paramagnetic channel of the Higgs mode
having almost no polarization-angle dependence,
while the second leading contribution comes from the paramagnetic channel of quasiparticles generally showing significant polarization-angle dependence.
The result is consistent with the recent experimental observation of no polarization-angle dependence of THG,
giving firm evidence that the Higgs mode dominantly contributes to the THG resonance in NbN superconductors. 
\end{abstract}


\date{\today}

\maketitle

\section{Introduction}
\label{intro}

The standard microscopic theory of superconductivity, i.e., the BCS theory, predicts the presence of the collective amplitude mode
of the superconducting order parameter \cite{Anderson1958, Schmid1968, VolkovKogan1973, Kulik1981, 
LittlewoodVarma1981, LittlewoodVarma1982}, which is recently referred to as
the Higgs mode due to the close analogy
with the Higgs boson in particle physics (for recent reviews, see \cite{PekkerVarma2015, ShimanoTsuji2019}). 
Despite the fundamental and universal aspects of the Higgs mode,
its observation in ordinary superconductors had been elusive until recently.
One exception was a superconductor $2H$-NbSe$_2$, which is special in the sense 
that superconductivity and charge density wave (CDW) coexist in a single material.
In this particular situation, the Higgs mode becomes Raman active, and
has been observed in the early stage
by Raman experiments \cite{SooryakumarKlein1980, SooryakumarKlein1981}
(see also \cite{Measson2014, Grasset2018, Grasset2019} for recent studies). 
However, the Higgs mode itself should exist
irrespective of the presence of CDW, so that its observation in superconductors without any other orders
has been long awaited.

The difficulty in observing the Higgs mode in superconductors without other coexisting orders is that
the Higgs mode does not linearly couple to external electromagnetic fields, and that the energy of the Higgs mode,
which lies around the superconducting gap energy $2\Delta$,
is in the terahertz (THz) frequency range, for which an intense light source had been lacking for a long time.
The recent development of THz laser techniques, however, has made it possible to excite the Higgs mode directly
through the nonlinear light-Higgs coupling \cite{TsujiAoki2015}.
In fact, coherent oscillation of the superconducting order parameter with frequency $2\Delta$ after irradiation with a monocycle
THz pulse has been observed in a superconducting NbN \cite{Matsunaga2013}.
Subsequently, resonant enhancement of third harmonic generation (THG) at the condition of 
$2\Omega=2\Delta$ with $\Omega$ being the incident light frequency has been reported for NbN
using multicycle THz pulses \cite{Matsunaga2014}.

While all these measurements are consistent with the interpretation that the Higgs mode is excited by
THz laser excitations, it is not sufficient to confirm that the mode energy is $2\Delta$, since
the pair-breaking energy of quasiparticles is also equal to $2\Delta$. This forces one to distinguish 
the collective Higgs mode from individual excitations of quasiparticles
by properties other than the mode energy.

One way to discriminate them is to measure the polarization-angle dependence of the resonant THG
\cite{Cea2016}. According to the BCS mean-field calculation in the clean limit
for a single-band model, the quasiparticle contribution has strong angle dependence in THG,
whereas the Higgs-mode contribution does not. 
Followed by the theoretical proposal, the polarization-resolved measurement of THG has been performed for a single-crystal NbN,
showing that the THG intensity at the resonance has almost no polarization-angle dependence
\cite{Matsunaga2017}. 
Does this mean that the origin of the resonant THG observed in NbN is the Higgs mode?

The story is not so simple, because the BCS clean limit calculation
also suggests that the absolute magnitude of the quasiparticle contribution to the THG resonance is generally much larger than
that of the Higgs mode in the BCS clean limit \cite{Cea2016, ShimanoTsuji2019}. 
Considering both the polarization-angle dependence and absolute magnitude of the Higgs and quasiparticle contributions to the THG, we come to the conclusion
that at least the BCS mean-field treatment in the clean limit
fails to describe the THG experiments for NbN superconductors.


There are several possibilities to circumvent this 
controversial situation: 
One is to go beyond the BCS approximation and include, e.g., phonon retardation effects. 
In fact, NbN is known to have a moderately strong electron-phonon coupling (with a dimensionless
coupling constant $\lambda\sim 1$) \cite{Kihlstrom1985, Brorson1990, Chockalingam2008}.
Based on the nonequlibrium dynamical mean-field theory \cite{noneqDMFTreview},
it has been shown that the Higgs mode can contribute to THG with an order of magnitude comparable to
quasiparticles \cite{TsujiMurakamiAoki2016}.

Another possibility is to depart from the clean limit and consider the effect of 
disorders or impurity scattering.
Since the optical conductivity of NbN used in the THz laser experiments
agrees well \cite{Matsunaga2013, Matsunaga2014} 
with the Mattis-Bardeen form \cite{MattisBardeen1958}, 
the NbN samples are close to the dirty limit. 
The effect of impurity scattering on THG in BCS superconductors has been
studied in Refs.~\cite{Jujo2018, MurotaniShimano2019, Silaev2019}. Strikingly, in the dirty regime
the magnitude of the Higgs-mode contribution to THG can exceed by far that of quasiparticles.

\begin{figure}[t]
    \centering
    \includegraphics[width=8cm]{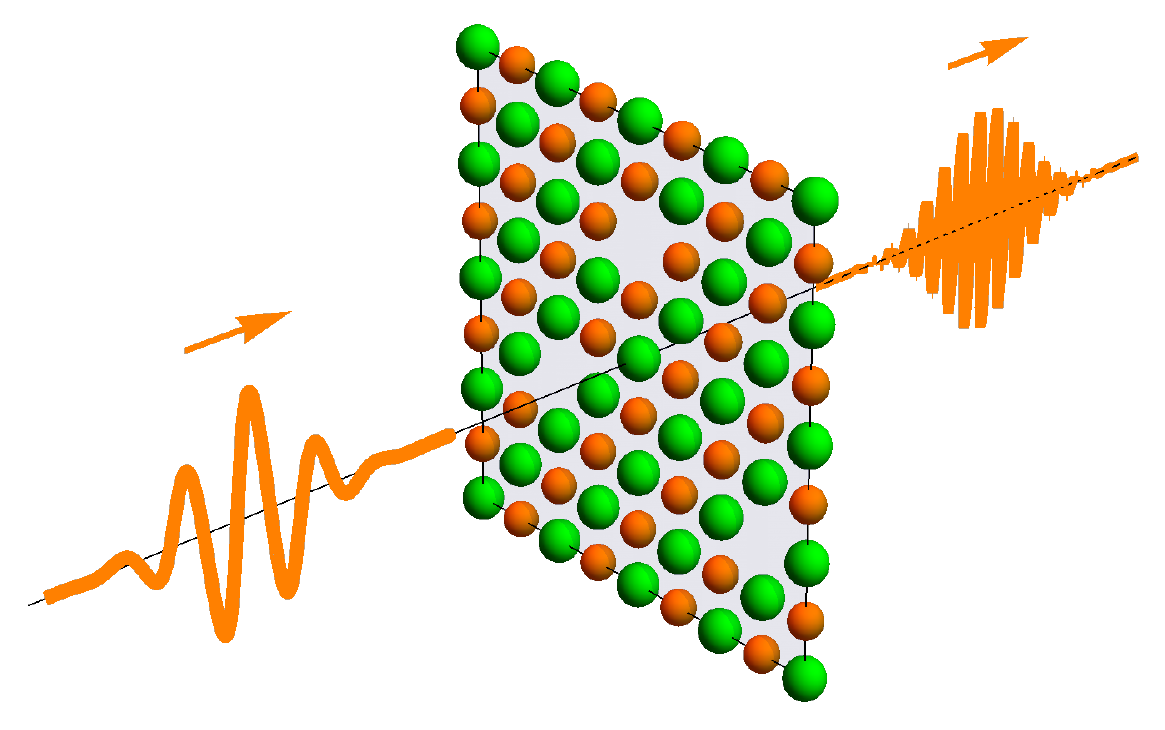}
    \caption{A schematic picture of third harmonic generation in 
    NbN superconductors with disorders or impurities 
    (for which we illustrate lattice defects as an example).}
    \label{fig:NbN}
\end{figure}

These studies suggest that impurity scattering has more substantial effects 
on the magnitude of the THG resonance than phonon retardation.
It is natural to expect so, because the impurity scattering rate generally approaches a nonzero constant value in the low-energy limit,
while the electron-phonon scattering rate decays to zero in the Fermi-liquid regime.
Hence, in the present study we focus on the effect of impurity scattering.
A crucial key to the open issue of which of the Higgs mode or quasiparticles are dominant in the THG resonance in NbN 
is the polarization-angle dependence of THG in the dirty regime of superconductors, 
which has not been addressed so far.

In this paper, we study the polarization-angle dependence of THG in NbN superconductors with disorders (Fig.~\ref{fig:NbN}). 
For this purpose, we use an effective three-band model including the phonon-mediated multiband pairing interactions for NbN. 
In particular, we take special care of the relative magnitude between the intraband and interband pairing interactions, 
since the polarization-angle dependence might be strongly affected by it.
In the previous study \cite{Matsunaga2017}, the calculations of the polarization-angle dependence for the three-band model have been performed at the BCS clean limit without first-principles estimate of the pairing interactions. 
The calculations assuming the same amplitude of the intraband and interband pairing interaction have shown that the Higgs-mode contribution in THG is isotropic,
while the quasiparticle contribution has significant angle dependence \cite{Matsunaga2017}. 
For other choices of the relative magnitude between the intraband and interband interaction parameters,
the Higgs mode can also exhibit the polarization-angle dependence \cite{Cea2018}. 
In the present study, we go beyond these previous studies 
by taking into account the effect of impurities with a realistic estimate of the ratio between the intraband and interband interactions.




We first estimate the pairing interaction parameters for NbN 
from first principles calculations of the phonon band structure and the electron-phonon couplings. 
Using the estimated ratio between the intraband and interband interaction parameters,
we calculate the THG susceptibility for the multiband superconductor NbN within the BCS mean-field theory. 
The effect of nonmagnetic impurity scattering is treated
by means of the self-consistent Born approximation. 
We consider both intraband and interband impurity scatterings for multiband NbN superconductors.
The calculated THG susceptibility is classified according to the physical origin (quasiparticle or Higgs mode), the coupling channel to light (diamagnetic or paramagnetic), and the diagrammatic representation in the presence of impurities.


The results show that the THG resonance is dominated by the paramagnetic channel in the dirty regime in NbN, in which
the Higgs-mode contribution generally becomes larger than the quasiparticle contribution. 
This behavior is similar to the previous results for single-band superconductors \cite{Jujo2018, Silaev2019}.
With the estimated ratio between the intraband and interband pairing interactions,
the quasiparticles always show clear polarization-angle dependence of THG, 
while the Higgs mode does not in general, except in the vicinity of the parameter region
where the interband impurity scattering rate vanishes.
By comparing with the polarization-resolved THG measurements for NbN \cite{Matsunaga2017}, 
we conclude that the dominant contribution to the THG resonance is coming from the Higgs mode
rather than quasiparticles.


The paper is organized as follows. In Sec.~\ref{el-ph coupling}, we evaluate the electron-phonon couplings
and the effective pairing interactions in NbN from first principles calculations. 
In Sec.~\ref{method}, we describe the method to calculate the THG susceptibility using an effective three-band model for NbN
with multiband pairing interactions and impurity scatterings.
In Sec.~\ref{THG}, we show the numerical results for THG in NbN superconductors,
focusing on its magnitude and polarization-angle dependence
of the Higgs-mode and quasiparticle contributions
for various impurity scattering rates. 
The paper is summarized in Sec.~\ref{summary}.

\section{First principles estimation of the electron-phonon coupling in ${\rm \bf NbN}$}
\label{el-ph coupling}

In this section, 
we evaluate the intraband and interband effective pairing interactions of NbN 
from first principles,
which are important to determine the polarization-angle dependence of THG in multiband NbN superconductors
as discussed in the introduction. The electronic band structure of NbN has been calculated from first principles in the previous literatures 
\cite{Mattheiss1972, FongCohen1972, ChadiCohen1974, Amriou2003, Matsunaga2017}.
The {\it ab initio} estimate of the phonon band structure and the electron-phonon coupling constant of NbN
has been reported in \cite{Papaconstantopoulos1985, Isaev2005, Isaev2007, Blackburn2011}.
While the total effective pairing interaction (summed over the band indices) has been derived in the previous calculations,
here we need the band-resolved matrix elements of the pairing interaction.

Our approach is based on the {\it ab initio} construction of a low-energy effective model of NbN including the electron-phonon coupling.
To this end, we perform the density functional calculation for NbN using \textsc{Quantum ESPRESSO} package \cite{Giannozzi2009, Giannozzi2017}.
We use the Troullier-Martins norm-conserving pseudopotentials~\cite{Troullier1991} in the Kleinman-Bylander representation~\cite{Kleinman1982} 
with the Perdew-Burke-Ernzerhof \cite{Perdew1996} exchange-correlation functional. 
We set the cutoff energy for the wave functions and charge density to be 100 eV and 400 eV, respectively, 
and take $N_{\bm k}=8\times 8\times 8$ $\bm k$ points for electron's momentum mesh.

The phonon band structure and the electron-phonon coupling constants are evaluated by the density functional perturbation theory \cite{Baroni2001},
for which we use $N_{\bm q}=8\times 8\times 8$ $\bm q$ points for phonon's momentum mesh.
The previous phonon band calculation \cite{Isaev2005, Isaev2007} shows that
NbN in the NaCl-type structure has a structural instability as indicated by imaginary phonon frequencies, which is, however, not observed in experiments. 
To avoid such an instability, we employ a virtual crystal approximation, where we create a pseudopotential for Nb with the nuclear charge $Z=+40.5$.
With this, we fully optimize the lattice structure, obtaining the lattice constant $a=4.497$ \AA, which agrees well with the experimental data \cite{Heger1980}.

\begin{figure}[t]
\includegraphics[width=8.5cm]{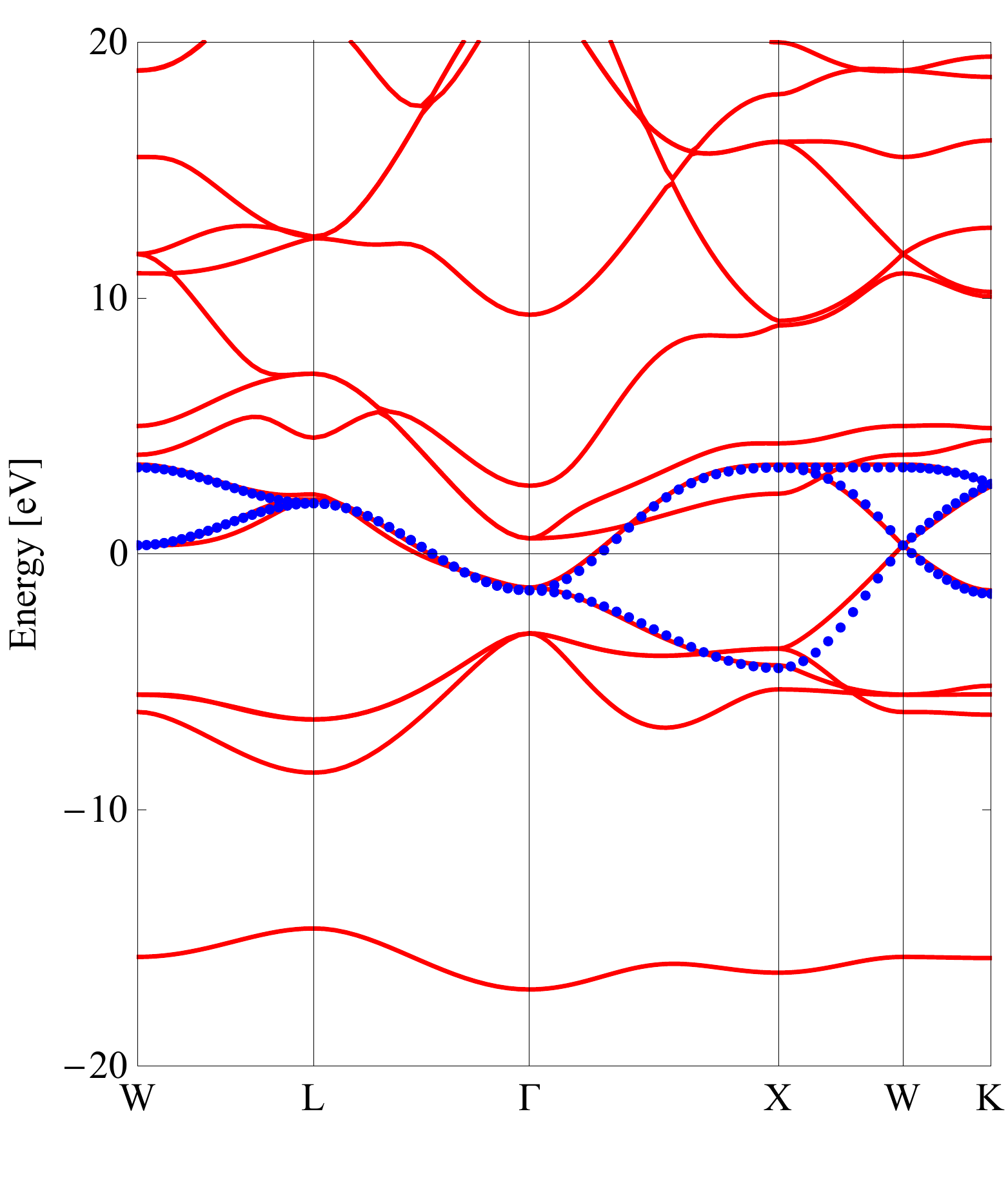}
\caption{The electron band dispersion of NbN obtained from the first-principles calculation (red curves) 
and that of the simplified model [Eq.~(\ref{simplified H}), blue dots] with the hopping parameters fitted to
the effective three-band model constructed from the maximally localized Wannier orbitals.}
\label{electron band}
\end{figure}

Our calculation of the electronic band structure of NbN (red curves in Fig.~\ref{electron band}) well reproduces the previous results
\cite{Mattheiss1972, FongCohen1972, ChadiCohen1974, Amriou2003, Matsunaga2017}.
Near the Fermi energy, there are three bands consisting of Nb's $4d$ $t_{2g}$ orbitals ($xy$, $yz$ and $zx$), 
which are occupied by two electrons
in one unit cell in average. Therefore, NbN can be effectively regarded as a three-band system at one third filling
at low energy.
We first construct the effective three-band tight-binding Hamiltonian on the basis of the maximally localized Wannier orbitals \cite{Marzari1997, Souza2001}, for which we use the open-source package RESPACK \cite{Nakamura2020}.
We can simplify the effective three-band model by taking the leading hopping processes \cite{Matsunaga2017},
\begin{align}
H_{\rm el}
&=
\sum_{\bm k n \sigma} \epsilon_{\bm k n} c_{\bm k n \sigma}^\dagger c_{\bm k n \sigma}
\quad
(n=xy, yz, zx), 
\label{Eq_H_el}
\end{align}
where $c_{\bm k n \sigma}^\dagger$ is a creation operator of electrons with momentum $\bm k$, orbital $n$, and spin $\sigma$.
The simplified energy dispersion for the $d_{xy}$ orbital is given by 
\begin{align}
\epsilon_{\bm k,xy}
&=
4t \cos \frac{k_x}{2} \cos \frac{k_y}{2}
+2t'(\cos k_x+\cos k_y)
\notag
\\
&\quad
+4t''\left(
\cos\frac{k_y}{2}\cos\frac{k_z}{2}+\cos\frac{k_z}{2}\cos\frac{k_x}{2}\right).
\label{simplified H}
\end{align}
The remaining band dispersions $\epsilon_{\bm k, yz}$ and $\epsilon_{\bm k, zx}$ are given by
permuting $x$, $y$, and $z$ in $\epsilon_{\bm k, xy}$.
The three hopping parameters $t$, $t'$, and $t''$ are fitted with the three bands
constructed from the maximally localized Wannier orbitals.
The results are $t=-0.79$ eV, $t'=-0.22$ eV, and $t''=0.19$ eV,
which slightly deviate from the previous result in Ref.~\cite{Matsunaga2017}.
The difference arises because the previous study fit the expression (\ref{simplified H}) directly with the original band structures 
(corresponding to red curves in Fig.~\ref{electron band}), while here we fit Eq.~(\ref{simplified H}) using the band dispersion of the maximally localized Wannier orbitals. 

\begin{figure}[t]
\includegraphics[width=8.5cm]{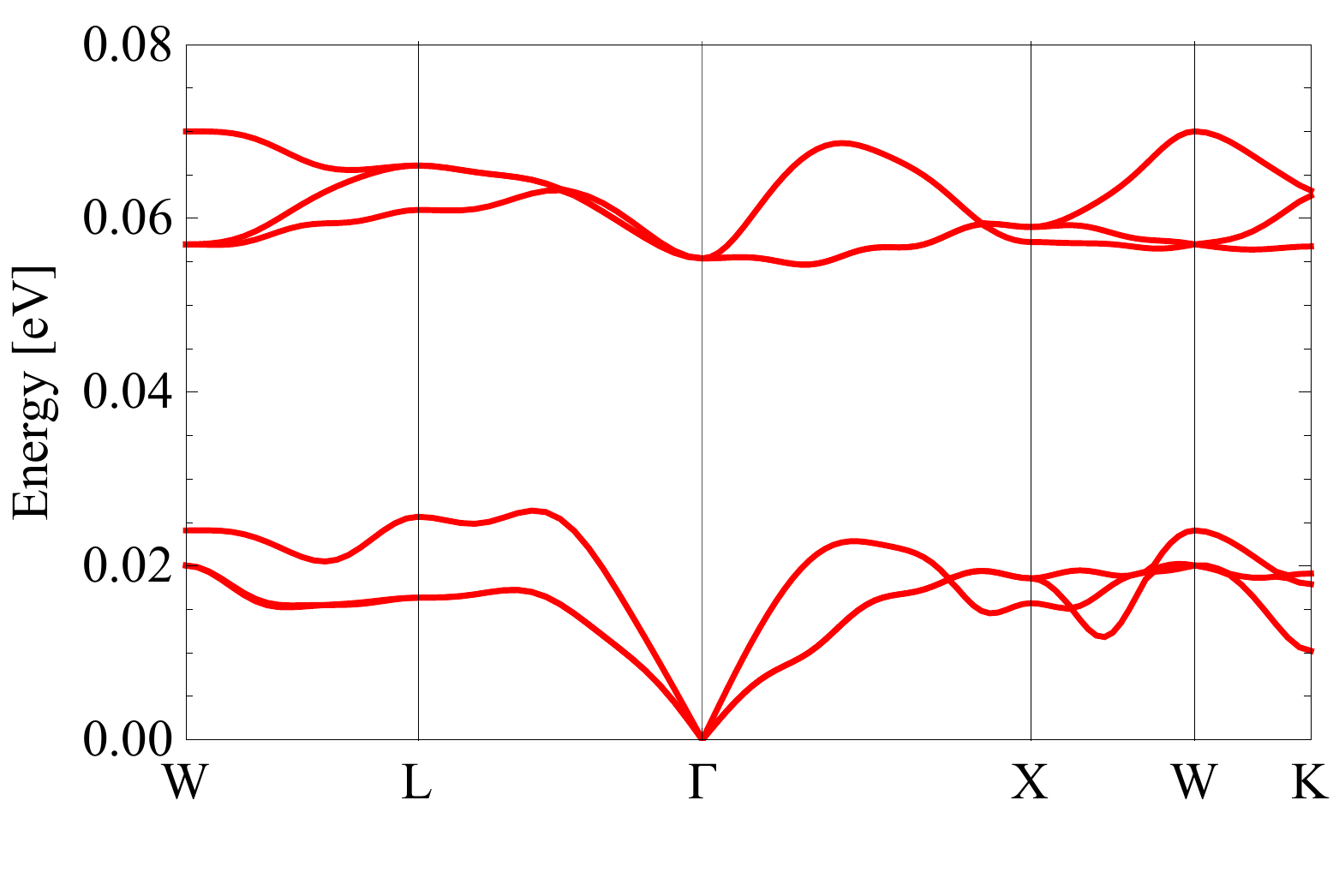}
\caption{The phonon dispersion of NbN obtained from the first principles calculation.}
\label{phonon band}
\end{figure}

In Fig.~\ref{electron band}, we show the simplified band dispersions $\epsilon_{\bm kn}$ by blue dots.
One can see that both of the band dispersions constructed from the density functional calculation and from the simplified model (\ref{simplified H})
agree fairly well with each other.
We employ the simplified dispersion $\epsilon_{\bm kn}$ for the model-based calculation of THG in Sec.~\ref{THG},
where we need higher-order derivatives of the dispersion such as $\frac{\partial^2\epsilon_{\bm kn}}{\partial k_i \partial k_j}$
that can be analytically evaluated with the expression (\ref{simplified H}).

\begin{table*}[t]
\caption{The magnitude of the static part ($\omega=0$) of the Fermi-surface(FS)-averaged phonon-mediated attractions estimated from first principles for NbN.
The column ``Broadening'' shows the broadening width $\eta$ of the delta functions used to evaluate the FS average.
 $V_{\rm intra}$ and $V_{\rm inter}$ denote the intraband and interband attractions, respectively (see the text for details). }
\label{lambda table}
\begin{center}
\begin{tabular}{cccc}
\hline
\hline
\hspace{.2cm} Broadening [Ry] \hspace{.5cm} &  \hspace{0.4cm} $\left\langle V_{\rm intra} \right \rangle_{\rm FS}$ [eV] \hspace{0.4cm}
& \hspace{0.4cm} $\left\langle V_{\rm inter} \right \rangle_{\rm FS}$ [eV] \hspace{0.4cm} & \hspace{.2cm} $\left\langle V_{\rm inter} \right \rangle_{\rm FS}/\left\langle V_{\rm intra} \right \rangle_{\rm FS}$ \hspace{.2cm}
\\
\hline
 0.005    &     6.797    &     1.213   &     0.178
\\
 0.010    &     6.191    &     1.099   &     0.178
\\
 0.015    &     6.236    &     1.069   &     0.171
\\
 0.020    &     6.411    &     1.068   &     0.167
\\
 0.025    &     6.528    &     1.072   &     0.164
\\
 0.030    &     6.543    &     1.078   &     0.165
\\
 0.035    &     6.488    &     1.079   &     0.166
\\
 0.040    &     6.421    &     1.073   &     0.167
\\
 0.045    &     6.362    &     1.063   &     0.167
\\
 0.050    &     6.313    &     1.052   &     0.167
\\
\hline
\hline
\end{tabular}
\end{center}
\end{table*}

The effective model for phonons is represented by the Hamiltonian,
\begin{align}
H_{\rm ph}
&=
\sum_{\bm q\nu} \omega_{\bm q \nu} b_{\bm q \nu}^\dagger b_{\bm q \nu},
\end{align}
where $\omega_{\bm q \nu}$ is the phonon frequency, and $b_{\bm q \nu}^\dagger$
is the creation operator of phonons with momentum $\bm q$ at $\nu$th branch.
There are six phonon modes ($\nu=1,\dots,6$) in total, corresponding to two atoms (Nb and N) in the unit cell
each of which can oscillate along three orthogonal directions ($x$, $y$, and $z$).
In Fig.~\ref{phonon band}, we plot the phonon band dispersion $\omega_{\bm q\nu}$ of NbN obtained from the first principles calculation.
Three of them are acoustic phonons with linear dispersions around $\Gamma$ point, while the rest are optical phonons with energy gaps. 
Here we do not see imaginary phonon frequencies, implying that the present lattice structure is dynamically stable.

The electron-phonon coupling term is written as
\begin{align}
H_{\rm el\mathchar`-ph}
&=
\frac{1}{\sqrt{N_{\bm q}}}\sum_{\bm k \bm q} \sum_{mn\nu\sigma} 
g_{mn}^\nu(\bm k,\bm q) c_{\bm k+\bm q m \sigma}^\dagger c_{\bm k n \sigma}
\notag
\\
&\quad\times
(b_{\bm q \nu} + b_{-\bm q \nu}^\dagger),
\label{el-ph}
\end{align}
where 
$m$, $n=$ $xy$, $yz$, $zx$, and 
$g_{mn}^\nu(\bm k, \bm q)$ represents the matrix elements of the multiband electron-phonon coupling constant
estimated from the density function perturbation theory.
The calculations of the matrix elements are done in the maximally localized Wannier orbital basis \cite{Nomura2014}\footnote{Note that our tight-binding model in Eq. (\ref{Eq_H_el}) is written in the Wannier basis. 
Since orbital-off-diagonal hoppings are negligible, the Wannier and band indices agree with each other.}. 

The electron-phonon coupling mediates an effective attraction between the electrons, 
$-V_{mn}(\bm k,\bm q, \omega)=\sum_{\nu} |g_{mn}^\nu(\bm k,\bm q)|^2 D^{\rm ph}_{{\bm q}\nu} (\omega)$,
where we factor out the minus sign in front of $V_{mn}$ to indicate the attractive interaction, and $D^{\rm ph}_{{\bm q}\nu} (\omega)=2\omega_{\bm q\nu}/(\omega^2-\omega_{\bm q\nu}^2)$ is the phonon propagator.
If we neglect the retardation effect of the phonon-mediated attraction and take the static part $(\omega=0)$,
the phonon-mediated attraction is given by 
$V_{mn}(\bm k,\bm q) =  \sum_{\nu} |g_{mn}^\nu(\bm k,\bm q)|^2 \frac{2}{\omega_{\bm q \nu}}$.
By taking the momentum average of the attraction, we obtain the following BCS-type Hamiltonian:
\begin{align}
H_{\rm el\mathchar`-el}
&= 
-\frac{1}{N_{\bm k}}\sum_{\bm k \bm k' m n} V_{mn} c_{\bm k m \uparrow}^\dagger c_{-\bm k m \downarrow}^\dagger c_{-\bm k' n \downarrow} c_{\bm k' n \uparrow}.
\label{eq-BCS}
\end{align}
Here, $V_{nn} \equiv V_{\rm intra}$ and $V_{mn} \equiv V_{\rm inter}$ ($m\neq n$) denote the averaged intraband and interband effective attractions, respectively.
To estimate the realistic ratio between $V_{\rm intra}$ and $V_{\rm inter}$, 
we compute the static part $(\omega=0)$ of the Fermi-surface(FS)-averaged phonon-mediated attraction as
\begin{align}
 \left\langle V_{mn} \right\rangle_{\rm FS}
&=\frac{1}{N_{\bm q}N_{\bm k}}
\sum_{\bm k \bm q \nu} \frac{w_{\bm k+\bm q m} w_{\bm k n}  }{D^2 (\epsilon_F)} |g_{mn}^\nu(\bm k, \bm q)|^2 \frac{2}{\omega_{\bm q \nu}}, \nonumber \\
\label{attraction}
\end{align}
where $D(\epsilon_F)$ is the density of states for each $t_{2g}$ orbital at the Fermi energy (by symmetry, the density of states is the same among $t_{2g}$ orbitals),
and $w_{\bm k n}$ is the weight of a maximally localized Wannier orbital $n$ at the Fermi energy at momentum ${\bm k}$ 
given by $w_{\bm k n}=\sum_{\alpha} |U^{\bm k}_{ n\alpha} |^2 \delta(\epsilon_{\alpha {\bm k} } - \epsilon_F)$. 
Here, $\alpha$ is the Kohn-Sham Bloch band index, 
and $U^{\bm k}_{ n\alpha}$ is the unitary matrix relating the Wannier and Bloch bases 
($c_{\bm k \alpha\sigma}^\dagger = \sum_n c_{\bm k n\sigma}^\dagger U^{\bm k}_{n\alpha}$). 
For the details of the derivation of Eq.~(\ref{attraction}), we refer to Appendix \ref{appendix: FS average}.
In practical numerical calculations, $w_{\bm k n}$ is calculated using the Gaussian smearing $e^{-x^2/2\eta^2}/\sqrt{2\pi} \eta $ with a broadening width $\eta$.


In Table~\ref{lambda table}, we list the magnitude of the static part of the FS-averaged phonon-mediated attraction $\left\langle V_{\rm intra} \right \rangle_{\rm FS}$ and $\left\langle V_{\rm inter} \right \rangle_{\rm FS}$ for NbN. 
We note that the density of states $D(\epsilon_F)$ is about 0.12 states/eV and that the coupling constant $\lambda = D(\epsilon_F) \left ( \left\langle V_{\rm intra} \right \rangle_{\rm FS} +2 \left\langle V_{\rm inter} \right \rangle_{\rm FS} \right)$ amounts to $\sim 1$, in accord with the previous estimates \cite{Papaconstantopoulos1985, Isaev2005, Isaev2007, Blackburn2011}. 
The important quantity in the following THG calculations is the ratio between the intraband and interband interactions $V_{\rm inter}/V_{\rm intra}$ in Eq. (\ref{eq-BCS}). 
We find that the first-principles estimate of the ratio $\left\langle V_{\rm inter} \right \rangle_{\rm FS}/\left\langle V_{\rm intra} \right \rangle_{\rm FS}$ is about 0.17-0.18. 
Referring to the {\it ab initio} value, in the following sections, we set the ratio $V_{\rm inter}/V_{\rm intra}$ to be 0.18.


\section{Method for the calculation of third harmonic generation}
\label{method}

Having evaluated the effective intraband and interband pairing interactions for NbN evaluated in the previous section,
we now move on to the calculation and classification of the THG susceptibility for NbN with impurities.
Here the nonlinear susceptibilities are defined by
expanding the current with respect to the amplitude of the external field,
\begin{align}
\bm e\cdot\bm j(t)
&=
\chi_1 A(t)+\chi_2 A(t)^2+\chi_3 A(t)^3+\cdots,
\end{align}
where $\bm e$ is the polarization vector along which the current is measured,
and $A(t)$ is the amplitude of the vector potential.
We are mostly concerned with the emitted light with $\bm e$ parallel to the incident light.
In parity symmetric systems (as is the case for NbN), the even-order terms are absent. The third coefficient $\chi_3$ is the THG susceptibility
that we are interested in.
The induced current is accompanied by the electric polarization, 
which couples to electromagnetic fields and emits light.
Thus, one can effectively regard the nonlinear susceptibilities as being proportional to
the amplitude of the emitted light.

Our method of evaluating the THG susceptibility is based on the BCS mean-field approximation,
where we neglect phonon retardation effects (see the discussion in Sec.~\ref{intro}). We also do not explicitly consider dynamical screening effects due to long-range Coulomb interactions, 
which do not significantly modify the behavior of THG in superconductors \cite{Cea2016}.
For the treatment of impurities,
we employ the self-consistent Born approximation, which is valid in the weak disorder case 
(i.e., the impurity scattering rate is much smaller than the hopping but can be larger than the superconducting gap).
In the calculation of the THG susceptibility, we need to take into account the vertex corrections
represented by impurity ladder diagrams \cite{AbrikosovBook, Jujo2018, Silaev2019}.

\subsection{Formalism}

Let us consider a multiband system described by the BCS Hamiltonian
with the intraband and interband pairing interactions and nonmagnetic impurity scatterings,
\begin{align}
H_{\rm BCS}(t)
&=
\sum_{\bm k n \sigma} \epsilon_{\bm k-\bm A(t), n} c_{\bm k n \sigma}^\dagger c_{\bm k n \sigma}
\notag
\\
&\quad
-\frac{1}{N_{\bm k}}\sum_{\bm k \bm k' m n} V_{mn} c_{\bm k m \uparrow}^\dagger c_{-\bm k m \downarrow}^\dagger
c_{-\bm k' n \downarrow} c_{\bm k' n \uparrow}
\notag
\\
&\quad
+\sum_{imn\sigma} v_{imn} (c_{im\sigma}^\dagger c_{in\sigma}+{\rm h.c.}),
\label{BCS Hamiltonian}
\end{align}
where $\bm A(t)$ is the vector potential for external electromagnetic fields,
and $v_{imn}$ represents the impurity potential that hybridizes $m$ and $n$ bands
at a lattice site $i$. We assume that $v_{imn}$ is a Gaussain random variable with the disorder average
given by $\langle v_{imn}v_{i'm'n'}\rangle_{\rm disorder}=\gamma_{mn}^2 \delta_{ii'}\delta_{mm'}\delta_{nn'}$.
Here the impurity scattering rate is parametrized by the intraband and interband ones,
$\gamma_{\rm intra}=\gamma_{nn}$ and $\gamma_{\rm inter}=\gamma_{mn}$ ($m\neq n$), respectively.

To calculate real-frequency spectra,
we introduce nonequilibrium (retarded, advanced, and lesser) Green's functions for multiband superconductors,
\begin{align}
G_{\bm kn,ab}^R(t,t')
&=
-i\theta(t-t')\langle \{\Psi_{\bm kn,a}(t), \Psi_{\bm kn,b}^\dagger(t')\}\rangle,
\\
G_{\bm kn,ab}^A(t,t')
&=
i\theta(t'-t)\langle \{\Psi_{\bm kn,a}(t), \Psi_{\bm kn,b}^\dagger(t')\}\rangle,
\\
G_{\bm kn,ab}^<(t,t')
&=
i\langle \Psi_{\bm kn,b}^\dagger(t')\Psi_{\bm kn,a}(t) \rangle,
\end{align}
where $\Psi_{\bm kn}^\dagger=(c_{\bm kn\uparrow}^\dagger \; c_{-\bm kn\downarrow})$
is the two-component Nambu spinor, $a,b=1,2$ are the Nambu space indices,
and $\theta(t)$ is the step function [$\theta(t)=1$ ($t\ge 0$)
and $\theta(t)=0$ ($t<0$)]. 
The Green's functions satisfy the following Dyson equation,
\begin{align}
&
[i\partial_t-\xi_{\bm k-\bm A(t)\hat\tau_3,n}\hat\tau_3]
\hat G_{\bm kn}^\alpha(t,t')
\notag
\\
&
-\int d\bar t \;
[\hat\Sigma_n(t,\bar t)\hat G_{\bm kn}(\bar t,t')]^\alpha
=
\delta^\alpha(t,t')
\quad
(\alpha=R,A,<),
\label{Dyson}
\end{align}
where $\xi_{\bm k,n}=\epsilon_{\bm k,n}-\mu$ 
($\mu$ is the chemical potential),
$\hat\tau_i$ ($i=1,2,3$) are Pauli matrices in the Nambu space, 
and $\hat\Sigma_n(t,t')$ is the self-energy.
We put a hat on a matrix which has Nambu spinor indices.
The superscript $\alpha$ should be understood according to the Langreth rule \cite{Langreth1976},
i.e., $(XY)^R=X^RY^R$, $(XY)^A=X^AY^A$, $(XY)^<=X^RY^<+X^<Y^A$, and so on. We use a convention of $\delta^R(t,t')=\delta^A(t,t')=\delta(t,t')$
(Dirac's delta function) and $\delta^<(t,t')=0$.
In equilibrium with $\bm A=0$, the retarded Green's function is given
in a Fourier transformed form as
\begin{align}
\hat G_{\bm kn}^R(\omega)
&=
\left[(\omega+i\epsilon)\hat\tau_0-\xi_{\bm kn}\hat\tau_3-\hat\Sigma_n^R(\omega)\right]^{-1},
\label{eq Dyson}
\end{align}
where $\epsilon$ is a positive infinitesimal constant
and $\hat\tau_0$ represents the unit matrix.
In equilibrium,
the advanced and lesser Green's functions are given by
$\hat G_{\bm kn}^A(\omega)=\hat G_{\bm kn}^R(\omega)^\dagger$
and $\hat G_{\bm kn}^<(\omega)=f(\omega)[\hat G_{\bm kn}^A(\omega)-\hat G_{\bm kn}^R(\omega)]$,
where $f(\omega)=1/(e^{\beta\omega}+1)$ is the Fermi distribution function
and $\beta=(k_BT)^{-1}$ is the inverse temperature.

\begin{figure}[htbp]
\includegraphics[width=8.5cm]{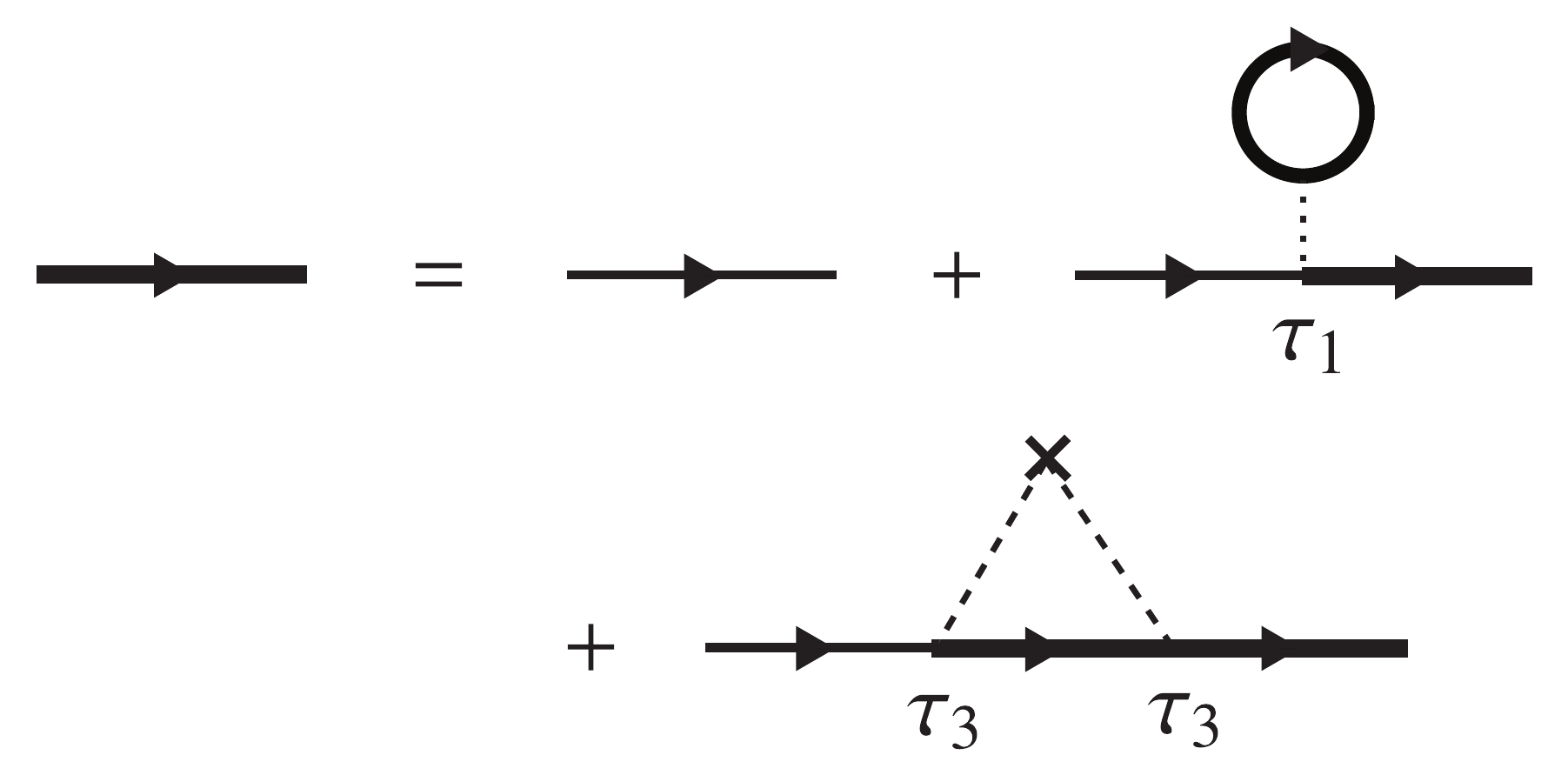}
\caption{The Dyson equation for the electron Green's function in the BCS mean-field and self-consistent Born approximations.
The thin and bold lines represent the noninteracting and interacting Green's functions, respectively.
The dotted line represents the pairing interaction, and the dashed lines represent the impurity scattering.
}
\label{self-consistent Born}
\end{figure}

In the BCS and self-consistent Born approximations, 
the self-energy is determined by
\begin{align}
\hat\Sigma_n^\alpha(t,t')
&=
\Delta_n(t)\hat\tau_1\delta^\alpha(t,t')
\notag
\\
&\quad
+\sum_m \gamma_{nm}^2 \frac{1}{N_{\bm k}}\sum_{\bm k}\hat\tau_3 \hat G_{\bm km}^\alpha(t,t')\hat\tau_3,
\label{retarded self-energy}
\end{align}
where $\Delta_n$ is the superconducting gap for a band $n$ defined by
\begin{align}
\Delta_n(t)
&=
\frac{i}{2}\sum_m V_{nm} \frac{1}{N_{\bm k}}\sum_{\bm k}
{\rm Tr}[\hat\tau_1 \hat G_{\bm km}^<(t,t)].
\label{gap}
\end{align}
In the above, we have assumed that Cooper pairs are formed within each band.
The equilibrium superconducting gap, self-energy, and Green's functions are self-consistently determined 
by Eqs.~(\ref{Dyson}), (\ref{retarded self-energy}), and (\ref{gap}).
In Fig.~\ref{self-consistent Born}, we show the diagrammatic representation for the Dyson equation
in the BCS mean-field and self-consistent Born approximations.

In order to evaluate the third harmonic generation, we employ the field-derivative approach developed in Ref.~\cite{TsujiMurakamiAoki2016},
which allows one to systematically derive nonlinear optical susceptibilities. The idea is to analytically differentiate the current,
\begin{align}
\bm j(t)
&=
-\frac{i}{N_{\bm k}}\sum_{\bm kn}
{\rm Tr} (\bm v_{\bm k-\bm A(t)\hat\tau_3,n} \hat G_{\bm kn}^<(t,t)),
\end{align}
with respect to the amplitude of the external field $\bm A(t)=\bm e Ae^{-i\Omega t}$,
where $\bm v_{\bm k,n}=\frac{\partial \epsilon_{\bm kn}}{\partial \bm k}$ is the group velocity,
$\bm e$ is the unit polarization vector ($\|\bm e\|=1$), $A$ and $\Omega$ are the amplitude and frequency of the field.
In this process, we repeatedly differentiate 
the self-consistent equations
(\ref{Dyson}), (\ref{retarded self-energy}), and (\ref{gap})
in the presence of the external field $\bm A(t)$.

Since odd-order derivatives of the self-energy $\hat\Sigma_n$ 
and the superconducting gap $\Delta_n$ vanish due to the parity symmetry of the system, what we need to calculate
for the THG susceptibility is the second derivatives,
$\ddot {\hat \Sigma}_n(\omega)=\frac{\partial^2}{\partial A^2}\hat\Sigma_n(\omega)$ 
and $\ddot{\Delta}_n=\frac{\partial^2}{\partial A^2}\Delta_n$ \cite{TsujiMurakamiAoki2016} 
(the derivative with respect to $A$ is denoted by dots), which are
self-consistently determined by the doubly differentiated equations.
By taking the second derivative of Eq.~(\ref{Dyson}), one obtains the relation
\begin{align}
\ddot{\hat G}_{\bm kn}^\alpha(\omega)
&=
[\hat G_{\bm kn}(\omega+2\Omega)
\ddot\epsilon_{\bm kn}\hat\tau_3\hat G_{\bm kn}(\omega)]^\alpha
\notag
\\
&\quad
+2[\hat G_{\bm kn}(\omega+2\Omega)
\dot\epsilon_{\bm kn}\hat G_{\bm kn}(\omega+\Omega)
\dot\epsilon_{\bm kn}\hat G_{\bm kn}(\omega)]^\alpha
\notag
\\
&\quad
+[\hat G_{\bm kn}(\omega+2\Omega)
\ddot{\hat\Sigma}_n(\omega)\hat G_{\bm kn}(\omega)]^\alpha
\label{ddot G}
\end{align}
with $\alpha=R,A,<$. 
In Eq.~(\ref{ddot G}), one can see that there are two types of couplings to the light field: 
one is the diamagnetic coupling ($\propto\rho\bm A^2$; $\rho$ is the electron density) given through $\ddot\epsilon_{\bm k}$, 
and the other is the paramagnetic coupling ($\propto\bm j\cdot \bm A$)
given through two $\dot\epsilon_{\bm k}$'s. The role of the latter paramagnetic coupling
has been emphasized as a dominant interaction between the Higgs mode and electromagnetic fields
\cite{TsujiMurakamiAoki2016}.

\begin{table*}[t]
\caption{Classification of the THG susceptibility according to the physical origin (quasiparticle or Higgs mode),
the coupling channel to light (diamagnetic or paramagnetic),
and the diagrammatic representation in the presence of impurities.
In the diagrams, the lines with arrows, single wavy lines, double wavy lines, and shaded squares represent 
the electron propagators, photon propagators, Higgs-mode propagators, and impurity ladder corrections, respectively. 
In each diagram, three of the four photon propagators carry incoming frequencies $\Omega$,
while the rest carries outgoing frequency $3\Omega$.
The fifth column shows whether each THG susceptibility has a resonance at frequency $2\Omega=2\Delta$.
The sixth column shows whether each THG susceptibility is robust against nonmagnetic impurities.}
\label{THG diagram}
\begin{center}
\begin{tabular}{cccccc}
\hline
\hline
susceptibility & \hspace{1cm} origin \hspace{1cm} & \hspace{.5cm} channel \hspace{.5cm} 
& diagram & resonance at $2\Omega=2\Delta$ & impurity robustness
\\
\hline
$\chi_{\rm qp}^{(1)}$ & quasiparticle & diamagnetic &
\begin{minipage}{2cm}
\vspace{.1cm}
\centering
\scalebox{0.55}{\includegraphics{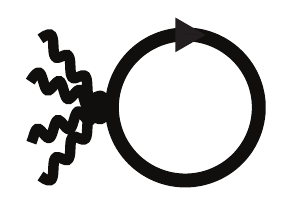}}
\end{minipage}
& & \checkmark
\\
$\chi_{\rm qp}^{(2)}$ & quasiparticle & paramagnetic &
\begin{minipage}{2cm}
\vspace{.1cm}
\centering
\scalebox{0.6}{\includegraphics{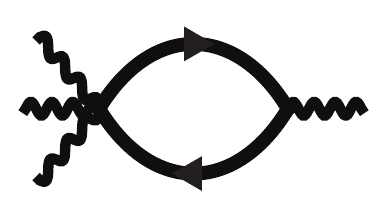}}
\end{minipage}
& & 
\\
$\chi_{\rm qp}^{(3)}$ & quasiparticle & diamagnetic &
\begin{minipage}{2cm}
\vspace{.1cm}
\centering
\scalebox{0.6}{\includegraphics{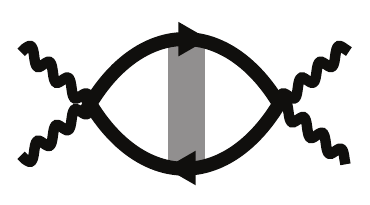}}
\end{minipage}
& \checkmark & \checkmark
\\
$\chi_{\rm qp}^{(4)}$ & quasiparticle & mixed &
\begin{minipage}{2cm}
\vspace{.1cm}
\centering
\scalebox{0.6}{\includegraphics{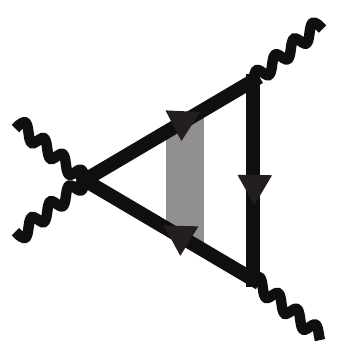}}
\end{minipage}
& \checkmark & 
\\
$\chi_{\rm qp}^{(5)}$ & quasiparticle & paramagnetic &
\begin{minipage}{2cm}
\vspace{.1cm}
\centering
\scalebox{0.6}{\includegraphics{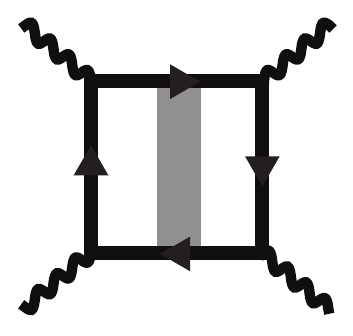}}
\end{minipage}
& \checkmark & 
\\
$\chi_{\rm H}^{(1)}$ & Higgs mode & diamagnetic &
\begin{minipage}{3.8cm}
\vspace{.1cm}
\centering
\scalebox{0.6}{\includegraphics{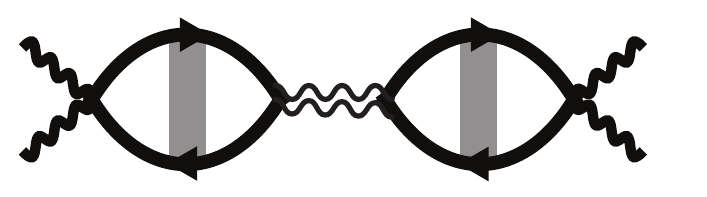}}
\end{minipage}
& \checkmark & \checkmark
\\
$\chi_{\rm H}^{(2)}$ & Higgs mode & mixed &
\begin{minipage}{3.8cm}
\vspace{.1cm}
\centering
\scalebox{0.6}{\includegraphics{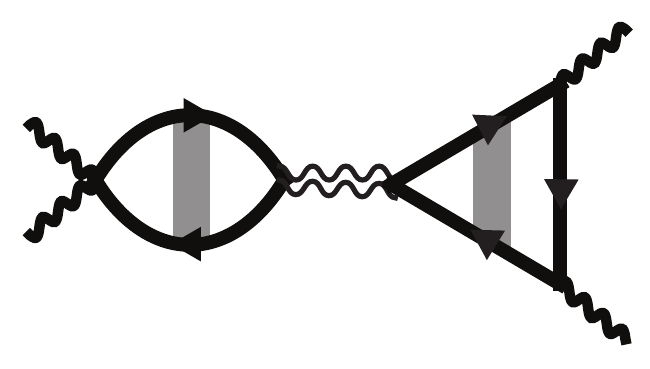}}
\end{minipage}
& \checkmark & 
\\
$\chi_{\rm H}^{(3)}$ & Higgs mode & paramagnetic &
\begin{minipage}{3.8cm}
\vspace{.1cm}
\centering
\scalebox{0.6}{\includegraphics{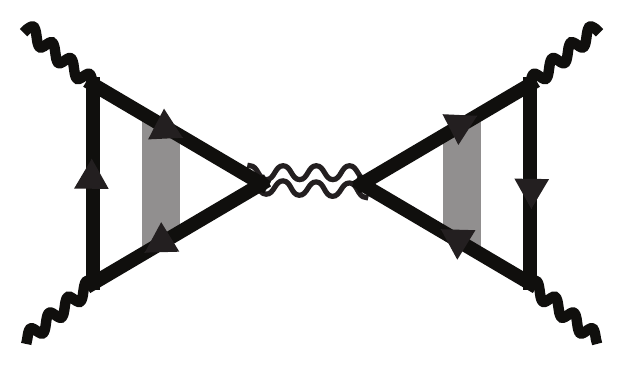}}
\end{minipage}
& \checkmark & 
\vspace{.1cm}
\\
\hline
\hline
\end{tabular}
\end{center}
\end{table*}

The second derivative of the self-energy $\ddot{\hat\Sigma}_n(\omega)$ is determined by
doubly differentiating Eq.~(\ref{retarded self-energy}),
\begin{align}
\ddot{\hat\Sigma}_n^\alpha(\omega)
&=
\ddot\Delta_n \hat\tau_1^\alpha
+\sum_m \gamma_{nm}^2 \frac{1}{N_{\bm k}}\sum_{\bm k}\hat\tau_3 \ddot{\hat G}_{\bm km}^\alpha(\omega)\hat\tau_3,
\label{ddot self-energy}
\end{align}
where $\tau_1^\alpha=\tau_1$ for $\alpha=R,A$ and $\tau_1^\alpha=0$ for $\alpha=<$.
The first term on the right hand side of Eq.~(\ref{ddot self-energy}) represents
the effect of amplitude fluctuation of the superconducting gap (i.e., the Higgs mode),
while the second term corresponds to the impurity-ladder vertex corrections.
Finally, the second derivative of the superconducting gap $\ddot\Delta_n$ is given by doubly differentiating
Eq.~(\ref{gap}),
\begin{align}
\ddot\Delta_n
&=
\frac{i}{2}\sum_m V_{nm} \frac{1}{N_{\bm k}}\sum_{\bm k}
\int \frac{d\omega}{2\pi} {\rm Tr}[\hat\tau_1 \ddot{\hat G}_{\bm km}^<(\omega)].
\label{ddot gap}
\end{align}

The second derivatives, $\ddot{\hat G}_{\bm kn}(\omega)$, $\ddot{\hat\Sigma}_n(\omega)$, and $\ddot\Delta_n$,
are calculated by solving the self-consistent equations (\ref{ddot G}), (\ref{ddot self-energy}), and (\ref{ddot gap}).
The results are plugged into the third derivative of the current $\dddot{\bm j}$ to obtain the THG susceptibility. 
In this paper, we focus on the THG induced along the polarization direction of the incident light.
More details of the derivation of the THG susceptibility are described in Appendix~\ref{appendix: THG}.

\subsection{Classification of THG susceptibilities}

\begin{figure}[htbp]
\includegraphics[width=8cm]{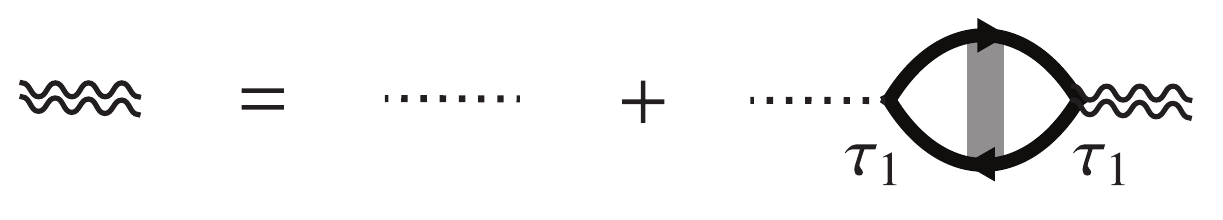}
\caption{The self-consistent equation for the Higgs-mode propagator (double wavy lines).}
\label{Higgs-mode propagator}
\end{figure}

\begin{figure}[htbp]
\includegraphics[width=8cm]{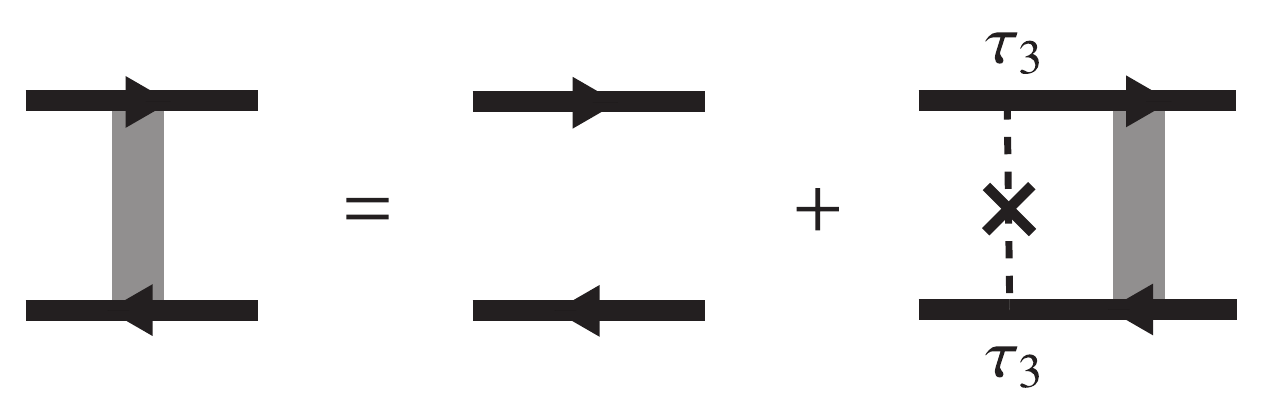}
\caption{The self-consistent equation for the impurity ladder correction (shaded squares).}
\label{impurity ladder}
\end{figure}

In Table~\ref{THG diagram}, we list all the THG diagrams including both 
the quasiparticle and Higgs-mode contributions.
The distinction between the two contributions is defined by 
whether or not the diagram includes the Higgs-mode propagator depicted by the double wavy lines. 
The Higgs-mode propagator contains the fluctuation of the superconducting gap amplitude,
which is diagrammatically shown in Fig.~\ref{Higgs-mode propagator}.
Here the dotted lines represent the bare attractive interaction $V_{mn}$,
while the bold lines with arrows represent the electron propagator $\hat G_{\bm kn}$.
The shaded square represents the impurity-ladder correction, as shown in Fig.~\ref{impurity ladder}.

There are five topologically inequivalent diagrams for quasiparticles and three diagrams for the Higgs mode.
They have different couplings to external laser fields (single wavy lines)
classified into the paramagnetic, diamagnetic, and mixed channels in Table~\ref{THG diagram}.
Each outer vertex attached to $\ell$ photon lines in the THG diagram is assigned to 
the $\ell$th derivative $\frac{d^\ell\epsilon_{\bm k}}{dA^\ell}$, which has the same parity as the density
if $\ell$ is even and has the same parity as the current if $\ell$ is odd.
Hence we call the channel of the coupling to light diamagnetic when $\ell$ is even and paramagnetic
when $\ell$ is odd. There are THG diagrams in which the paramagnetic and diamagnetic couplings coexist,
which we refer to as the mixed channel.
We remark that the impurity correction is absent for vertices with odd number of photon lines since odd parity terms vanish after momentum summation.

In Table~\ref{THG diagram}, we also show which THG susceptibility has a resonance at frequency $2\Delta$.
As we will see in Sec.~\ref{THG}, $\chi_{\rm qp}^{(i)}$ ($i=3,4,5$) and $\chi_{\rm H}^{(i)}$ ($i=1,2,3$) generally
show the resonance. The resonance of $\chi_{\rm H}^{(i)}$ originates from the collective Higgs mode
whose energy gap corresponds to $2\Delta$. On the other hand, the quasiparticle contributions $\chi_{\rm qp}^{(i)}$
also exhibit the resonance at $2\Delta$, which is equal to the lowest pair-breaking energy.
The degeneracy of the resonance energy between the Higgs mode and quasiparticles 
forces us to distinguish them by properties other than the resonance frequency, as discussed in Sec.~\ref{intro}.

We also indicate in Table~\ref{THG diagram} which THG susceptibility is 
robust (insensitive) against nonmagnetic impurity scattering.
Generally the THG susceptibility in the diamagnetic coupling channels do not depend on the impurity scattering rate, 
while the paramagnetic and mixed channels exhibit strong impurity dependence,
which plays a key role in enhancing the Higgs-mode contribution in THG in dirty regimes.
This fact is related to Anderson's theorem \cite{Anderson1959}
(which states robustness of the superconducting gap against nonmagnetic impurity scattering in equilibrium $s$-wave superconductors), 
which can be generalized to robustness of the Higgs mode \cite{Jujo2015, Jujo2018}.
We confirm the robustness of each THG channel against impurity scattering by numerical simulations in Sec.~\ref{THG}.

\section{Third harmonic generation in ${\rm \bf NbN}$ superconductor}
\label{THG}

Based on the method described in Sec.~\ref{method}, we numerically
evaluate the THG susceptibilities for NbN superconductors.
We use the simplified band dispersion of NbN
(\ref{simplified H}) derived from the first principles calculation in Sec.~\ref{el-ph coupling}. 
Throughout this section, we use eV as the unit of energy,
and fix the ratio between the intraband and interband 
phonon-mediated interactions to be $V_{\rm inter}/V_{\rm intra}=0.18$ (Sec.~\ref{el-ph coupling}).
The absolute value of the interaction is chosen such that the superconducting gap is fixed.
For the numerical feasibility, we take a relatively large superconducting gap $2\Delta=0.8$
to maintain sufficiently high frequency and momentum resolution
(cf. the real gap size is in the order of few meV). 
This requires us to take $50\times 50\times 50=125000$ $k$-points.
We have checked that the results do not change qualitatively
as we vary the value of $2\Delta$ within our reach of numerical calculations.
The intraband and interband impurity scattering rates $\gamma_{\rm intra}$ and $\gamma_{\rm inter}$ 
are free parameters. In order for the self-consistent Born approximation to be valid
(which is the case in the experimental situation \cite{Matsunaga2014, Matsunaga2017}),
the impurity scattering rates should be sufficiently smaller than
the Fermi energy ($\gamma_{\rm intra}, \gamma_{\rm inter}\ll \epsilon_F$). 
Here we restrict ourselves to $\gamma_{\rm intra}/2\Delta, \gamma_{\rm inter}/2\Delta \le 2.5$ (cf. $\epsilon_F\sim 3-4$).
We first focus on the case of $\gamma_{\rm intra}=2\gamma_{\rm inter}=\gamma$ with $0\le\gamma/2\Delta\le 2.5$.
Then we scan the parameter space of ($\gamma_{\rm intra},\gamma_{\rm inter}$) with $0\le\gamma_{\rm intra}/2\Delta,
\gamma_{\rm inter}/2\Delta\le 2.5$.
We set the filling to be one third for NbN superconductor (two electrons in the three bands) 
and the inverse temperature $\beta=50$, which is sufficiently lower than the superconducting critical temperature.
In the numerical simulation, we take a finite value of the constant $\epsilon=0.01$ 
[which has been introduced as a positive infinitesimal in Eq.~(\ref{eq Dyson})].

\subsection{Channel-resolved THG intensity}

We first present the results for the THG intensity in NbN superconductors
in the channel resolved manner as classified in Table~\ref{THG diagram} in Sec.~\ref{method}.
In this subsection, the polarization direction $\bm e$ of light is set to be parallel to $x$ crystal axis.

\begin{figure}[t]
\includegraphics[width=8.5cm]{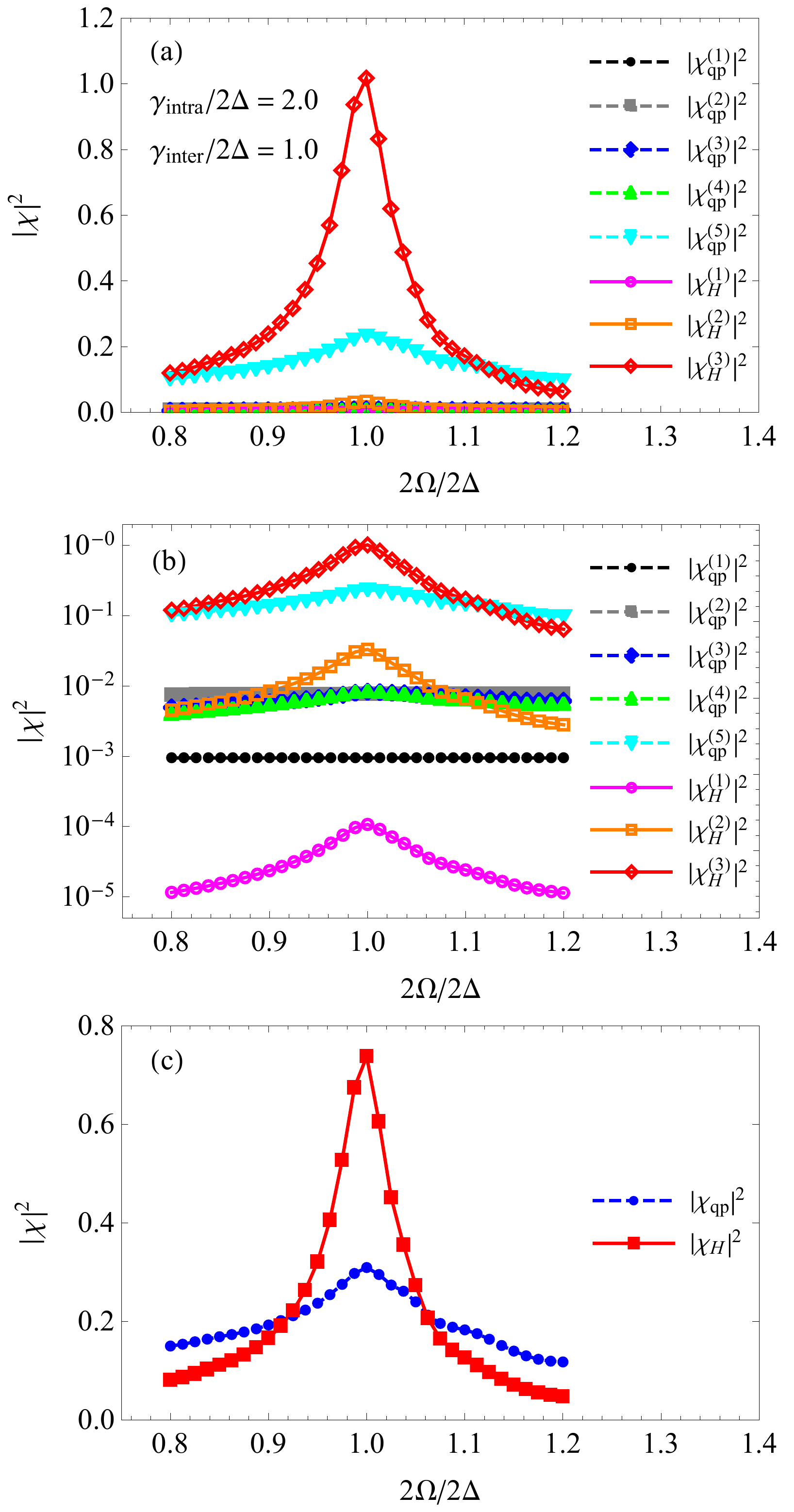}
\caption{(a), (b) Channel-resolved THG intensity for NbN superconductors
with $\gamma_{\rm intra}/2\Delta=2.0$ and $\gamma_{\rm inter}/2\Delta=1.0$
as a function of the frequency $2\Omega/2\Delta$
in the linear (a) and log scale (b).
(c) The total quasiparticle and Higgs-mode contributions to the THG intensity
for the model of NbN superconductors (\ref{BCS Hamiltonian}).
The polarization of light is parallel to $x$ crystal axis ($\theta=0^\circ$).}
\label{fig: THG omega}
\end{figure}

In Fig.~\ref{fig: THG omega}, we plot the frequency dependence of the THG intensity
for NbN superconductors with $\gamma_{\rm intra}/2\Delta=2.0$ and $\gamma_{\rm inter}/2\Delta=1.0$
(the dirty regime) in the linear [Fig.~\ref{fig: THG omega}(a)] and log scale [Fig.~\ref{fig: THG omega}(b)].
The leading contribution comes from the Higgs mode in the paramagnetic channel ($\chi_{\rm H}^{(3)}$),
showing a clear resonance peak at $2\Omega=2\Delta$.
The second dominant contribution comes from quasiparticles in the paramagnetic channel ($\chi_{\rm qp}^{(5)}$),
which has a relatively broadened resonance peak at $2\Omega=2\Delta$.
One can see that the resonance at $2\Omega=2\Delta$ occurs in the channels
$\chi_{\rm qp}^{(i)}$ ($i=3,4,5$) and $\chi_{\rm H}^{(i)}$ ($i=1,2,3$), being consistent with Table~\ref{THG diagram}.
In Fig.~\ref{fig: THG omega}(c), we plot the total quasiparticle
($|\chi_{\rm qp}|^2=|\sum_{i=1}^5 \chi_{\rm qp}^{(i)}|^2$) and Higgs-mode ($|\chi_{\rm H}|^2=|\sum_{i=1}^3\chi_{\rm H}^{(i)}|^2$) contributions
to the THG intensity in the dirty regime of NbN superconductors.
Clearly, the Higgs-mode contribution is larger than the quasiparticles.
This result agrees with the previous observations that the Higgs-mode contribution is drastically enhanced
due to impurity scattering \cite{Jujo2018, MurotaniShimano2019, Silaev2019}.
Similar enhancement has been found due to phonon retardation effects \cite{TsujiMurakamiAoki2016}.

\begin{figure}[t]
\includegraphics[width=8.5cm]{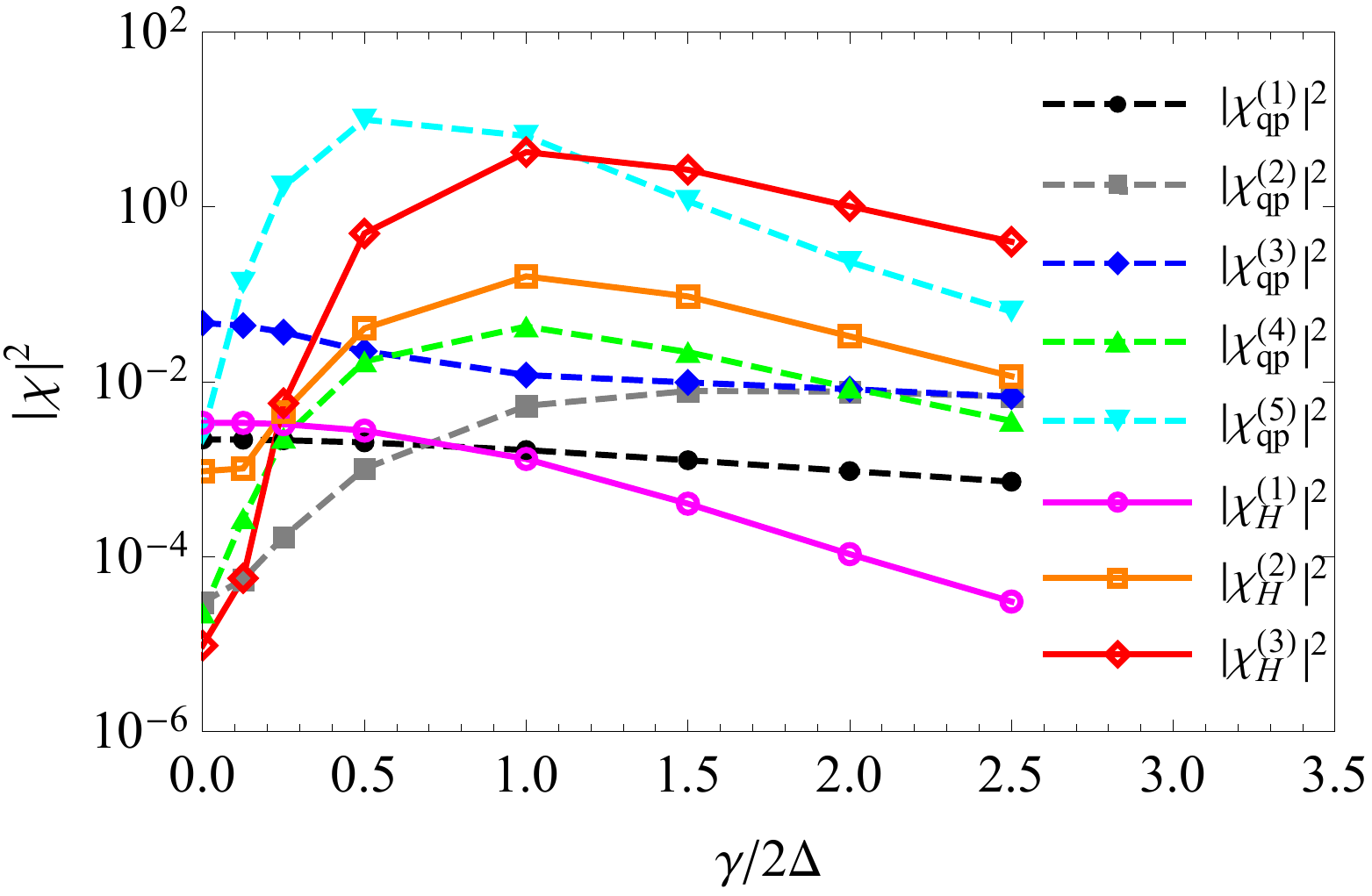}
\caption{Channel-resolved THG intensity for the model of NbN superconductors (\ref{BCS Hamiltonian}) at frequency $2\Omega=2\Delta$
as a function of the impurity scattering rate $\gamma_{\rm intra}=2\gamma_{\rm inter}=\gamma$.
The polarization of light is parallel to $x$ crystal axis ($\theta=0^\circ$).
The vertical axis is in the log scale.}
\label{fig: THG gamma}
\end{figure}

In Fig.~\ref{fig: THG gamma}, we plot the impurity dependence of 
the THG intensity for NbN superconductors at frequency $2\Omega=2\Delta$,
where we set $\gamma_{\rm intra}=2\gamma_{\rm inter}=\gamma$.
The THG intensity in the paramagnetic and mixed channels [$|\chi_{\rm qp}^{(i)}|^2$ ($i=2,4,5$) and
$|\chi_{\rm H}^{(i)}|^2$ ($i=2,3$)] show sensitive dependence on $\gamma$,
while those in the diamagnetic channel [$|\chi_{\rm qp}^{(i)}|^2$ ($i=1,3$) and $|\chi_{\rm H}^{(1)}|^2$]
is less sensitive (especially at small $\gamma/2\Delta$). 
This observation supports the general behavior of the THG susceptibility against impurities shown in Table \ref{THG diagram} in Sec.~\ref{method}.
The $\gamma$ dependence of the quasiparticle and Higgs-mode contributions in the diamagnetic and paramagnetic channels is qualitatively consistent with the previous order estimate in \cite{MurotaniShimano2019}.

\begin{figure}[htbp]
\includegraphics[width=8cm]{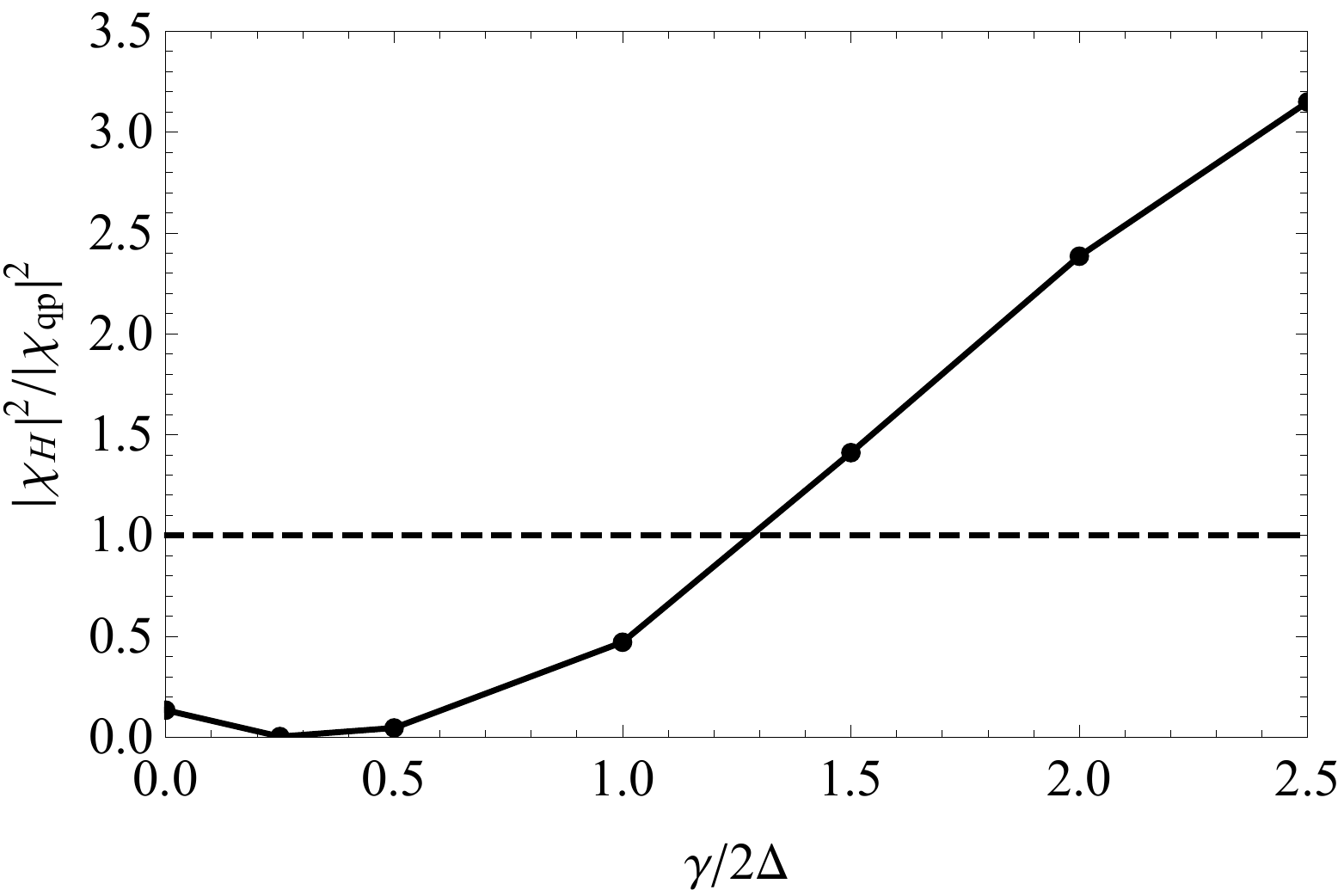}
\caption{The ratio $|\chi_{\rm H}|^2/|\chi_{\rm qp}|^2$ 
between the Higgs-mode and quasiparticle contributions to the THG intensity
for the model of NbN superconductors (\ref{BCS Hamiltonian})
as a function of the impurity scattering rate
$\gamma_{\rm intra}=2\gamma_{\rm inter}=\gamma$
at frequency $2\Omega=2\Delta$.
The polarization of light is parallel to $x$ crystal axis ($\theta=0^\circ$).}
\label{fig: THG ratio gamma}
\end{figure}

In the clean limit ($\gamma\to 0$), the most dominant contribution comes from
quasiparticles in the diamagnetic channel ($\chi_{\rm qp}^{(3)}$).
The second dominant one is the Higg-mode contribution in the diamagnetic channel ($\chi_{\rm H}^{(1)}$).
The paramagnetic channel is also present at $\gamma=0$, since we broaden the THG spectrum by taking 
the finite value of $\epsilon$ (so that the results at $\gamma=0$ slightly deviate from the ideal clean limit).
As we increase $\gamma$, the quasiparticle contribution in the paramagnetic channel ($\chi_{\rm qp}^{(5)}$) quickly grows,
and exceeds over the other components. At the same time, the Higgs-mode contribution
in the paramagnetic channel ($\chi_{\rm H}^{(3)}$) also grows rapidly.
Up to $\gamma/2\Delta\lesssim 1$, $\chi_{\rm qp}^{(5)}$ remains to be most dominant.
When the system enters the dirty regime ($\gamma/2\Delta \gtrsim 1$), 
the Higgs mode takes over the dominant part of the THG resonance, and $\chi_{\rm H}^{(3)}$ becomes the largest
contribution. 
This tendency seems to continue toward the dirty limit.
The maximum magnitude of the THG intensity in the paramagnetic channel is attained around $\gamma\sim \Delta$,
in agreement with the previous results \cite{MurotaniShimano2019}.

In Fig.~\ref{fig: THG ratio gamma}, we plot the ratio between 
the total Higgs-mode ($|\chi_{\rm H}|^2$) and quasiparticle ($|\chi_{\rm qp}|^2$) contributions to the THG intensity
for NbN superconductors
as a function of $\gamma_{\rm intra}=2\gamma_{\rm inter}=\gamma$.
In the clean regime ($\gamma\lesssim 2\Delta$) the quasiparticle contribution is dominant, 
whereas in the dirty regime ($\gamma\gtrsim 2\Delta$) the Higgs-mode contribution exceeds the quasiparticle one.
We expect that the ratio $|\chi_{\rm H}|^2/|\chi_{\rm qp}|^2$ continues to increase towards the dirty limit
(as observed in the single-band case \cite{Silaev2019}),
while our calculation is limited to $\gamma/2\Delta\le 2.5$ in order to maintain the validity of the self-consistent Born approximation.
One can see a little increase of $|\chi_{\rm H}|^2/|\chi_{\rm qp}|^2$ around $\gamma/2\Delta=0$,
which we attribute to the effect of the relatively large superconducting gap ($2\Delta=0.8$)
and the finite broadening factor ($\epsilon=0.01$) in our simulation.

\begin{figure}[t]
\includegraphics[width=8.5cm]{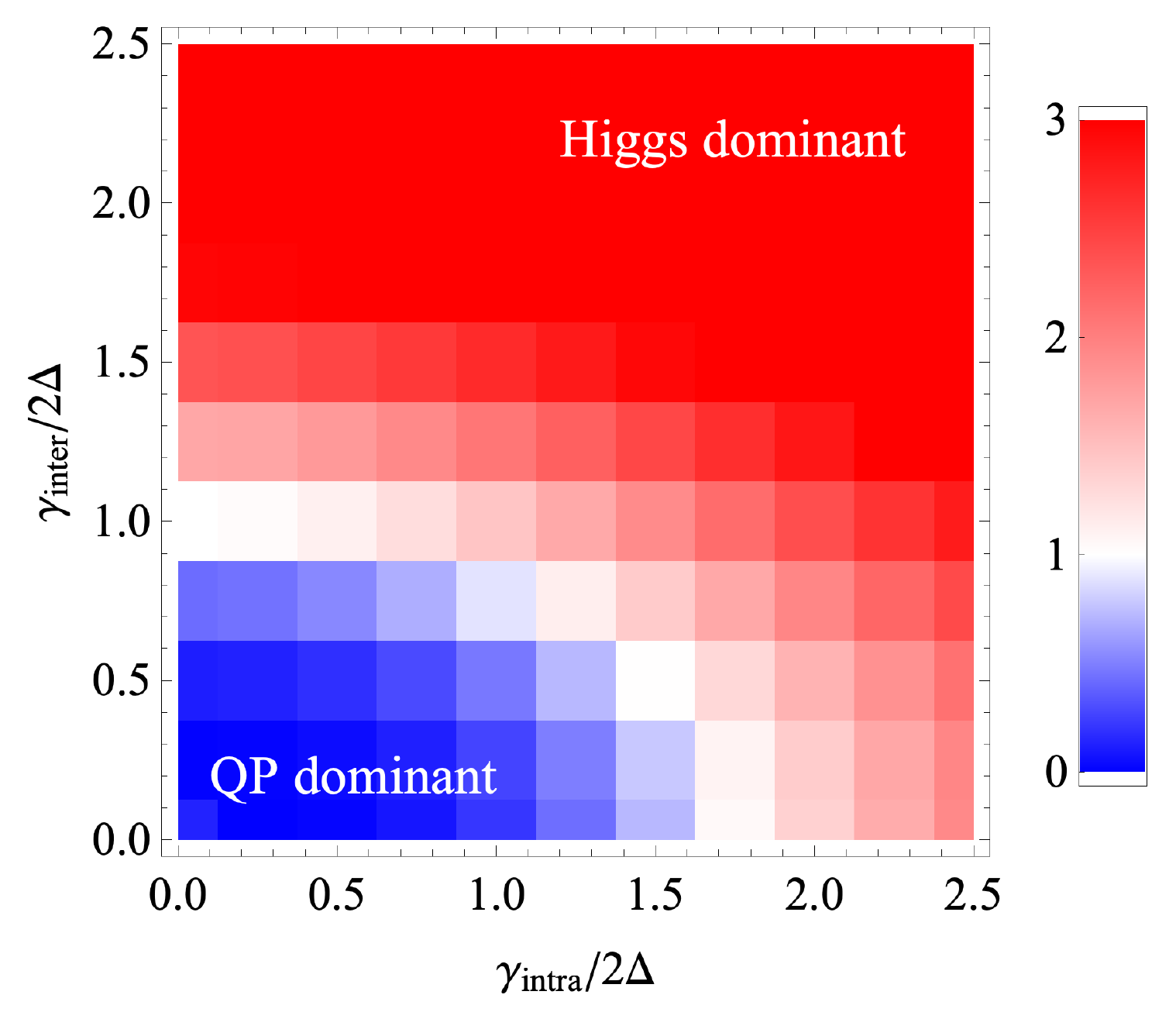}
\caption{The ratio $|\chi_{\rm H}|^2/|\chi_{\rm qp}|^2$ 
between the Higgs-mode and quasiparticle (QP) contributions to the THG intensity
for the model of NbN superconductors (\ref{BCS Hamiltonian}) at frequency $2\Omega=2\Delta$
plotted in the space of $\gamma_{\rm intra}/2\Delta$ and $\gamma_{\rm inter}/2\Delta$.
The polarization of light is parallel to $x$ crystal axis ($\theta=0^\circ$).}
\label{THG ratio}
\end{figure}

We separately change the intraband and interband impurity scattering rates to plot $|\chi_{\rm H}|^2/|\chi_{\rm qp}|^2$
in Fig.~\ref{THG ratio}. The quasiparticle dominant region is shown by blue in the color plot,
while the Higgs dominant region is shown by red. The boundary between the two regions is 
roughly given by $\gamma_{\rm intra}/2\Delta \sim 1.75$ and $\gamma_{\rm inter}/2\Delta \sim 1$.
The interband scattering is more effective to enhance the Higgs-mode contribution than 
the intraband one, since the interband scattering takes place more frequently in three-band systems.

\subsection{Polarization-angle dependence}
\label{polarization}

Next, we study the polarization-angle dependence of the THG intensity for NbN superconductors,
which may allow one to distinguish the quasiparticle and Higgs-mode contributions in experiments. 
The polarization angle is measured from the $x$ crystal axis, and the polarization vector $\bm e$ is rotated
in the $xy$ plane.

\begin{figure}[H]
\includegraphics[width=8.5cm]{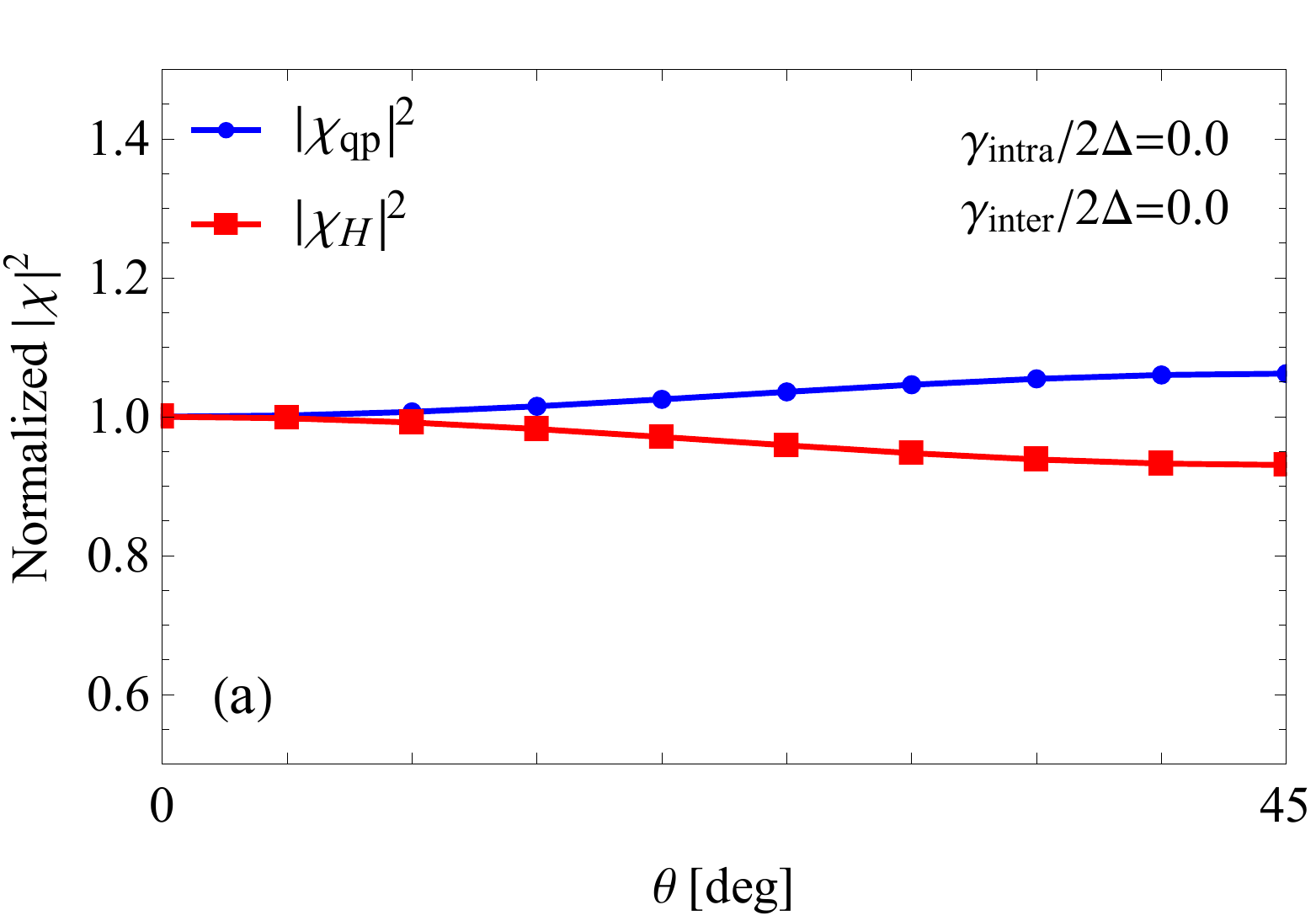}
\includegraphics[width=8.5cm]{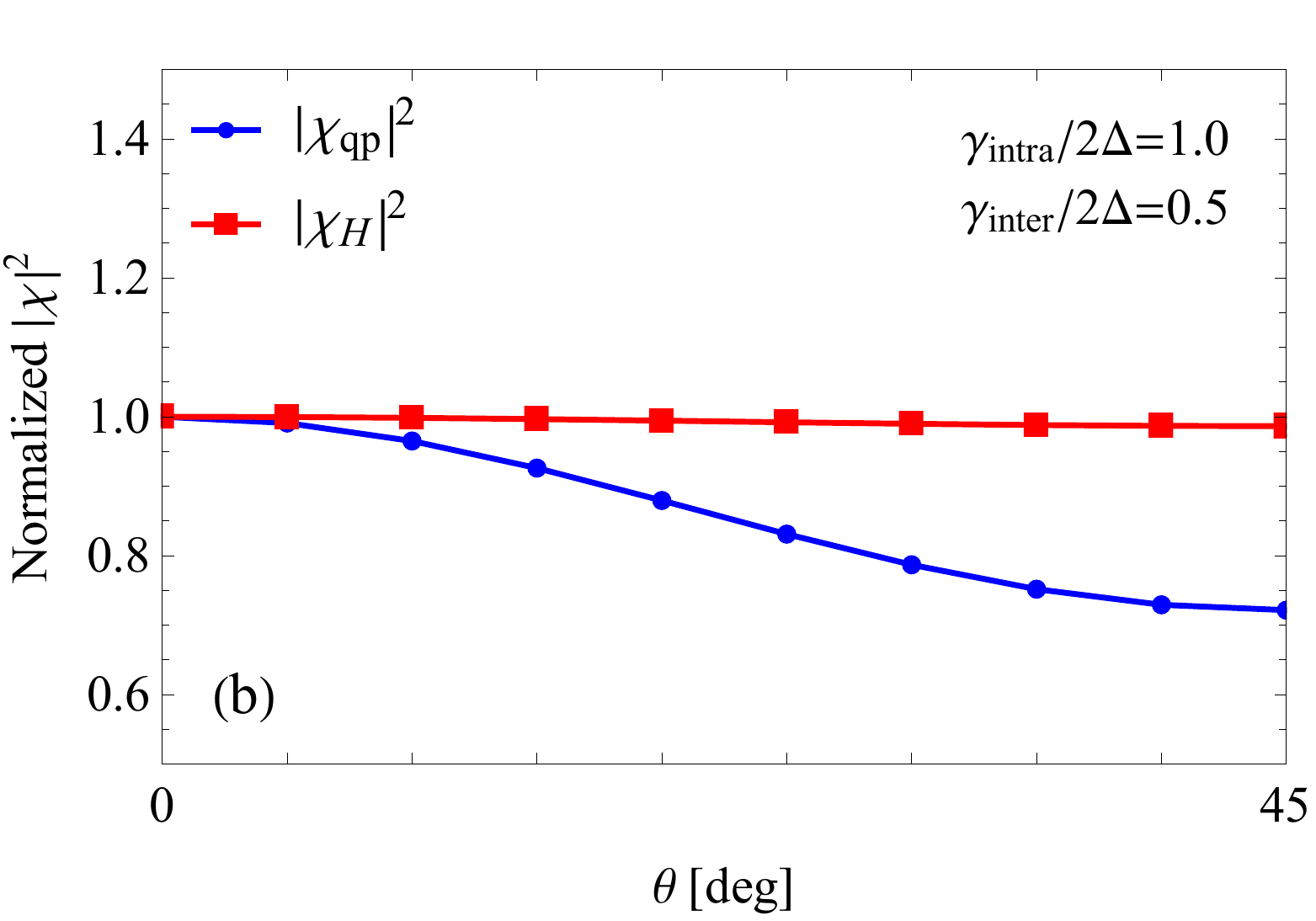}
\includegraphics[width=8.5cm]{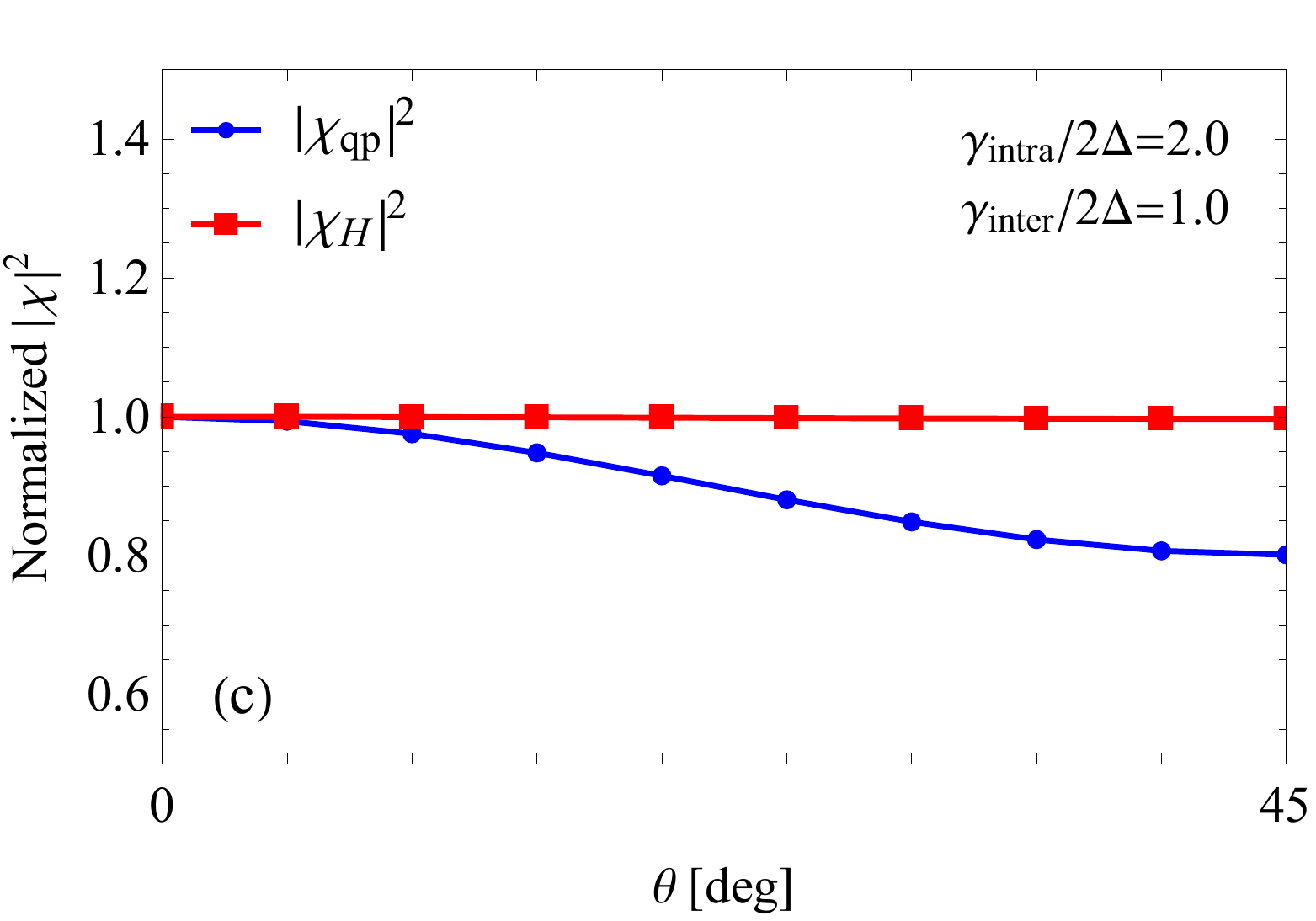}
\caption{The polarization-angle dependence of the quasiparticle and Higgs-mode contributions to the THG intensity 
for the model of NbN superconductors (\ref{BCS Hamiltonian})
at frequency $2\Omega=2\Delta$ with 
(a) $\gamma_{\rm intra}/2\Delta=0.0$, $\gamma_{\rm inter}/2\Delta=0.0$, 
(b) $\gamma_{\rm intra}/2\Delta=1.0$, $\gamma_{\rm inter}/2\Delta=0.5$, and 
(c) $\gamma_{\rm intra}/2\Delta=2.0$, $\gamma_{\rm inter}/2\Delta=1.0$.
Each quasiparticle and Higgs-mode contribution is normalized by the value at $\theta=0^\circ$, respectively.}
\label{fig: THG angle}
\end{figure}

In Fig.~\ref{fig: THG angle}, we plot the normalized $|\chi_{\rm qp}(\theta)|^2/|\chi_{\rm qp}(\theta=0^\circ)|^2$
and $|\chi_{\rm H}(\theta)|^2/|\chi_{\rm H}(\theta=0^\circ)|^2$ 
for several values of $\gamma_{\rm intra}/2\Delta$ and $\gamma_{\rm inter}/2\Delta$.
In the clean limit [Fig.~\ref{fig: THG angle}(a)], we find that 
the quasiparticle contribution grows monotonically by $\sim 6\%$ as the angle varies from $\theta=0^\circ$
to $45^\circ$, whereas the Higgs-mode contribution decreases by $\sim 7\%$.
The angle dependence of quasiparticles arises due to the anisotropic band structure of NbN.
The change of the quasiparticle contribution from $\theta=0^\circ$ to $45^\circ$ is smaller than 
that of the previous result \cite{Matsunaga2017}.
This is mainly due to the difference of the hopping parameters that we used in the effective model of NbN.
The angle dependence of the Higgs mode in the clean limit with small $V_{\rm inter}/V_{\rm intra}$
is consistent with the previous study \cite{Cea2018}.

As we increase $\gamma_{\rm intra}=2\gamma_{\rm inter}=\gamma$ 
[Fig.~\ref{fig: THG angle}(b),(c)], 
we observe qualitatively different polarization-angle dependence
for quasiparticles. Namely, $|\chi_{\rm qp}|^2$ decreases by $\sim 20-30\%$ 
as $\theta$ changes from $0^\circ$ to $45^\circ$.
This behavior is mostly coming from the paramagnetic channel, which becomes dominant in the dirty regime. Namely, the quasiparticle contribution in the paramagnetic channel always tends to decrease from $\theta=0^\circ$ to $45^\circ$ for arbitrary impurity scattering rates.
The transition from the increasing to decreasing dependence on the polarization angle is very rapid,
taking place around $\gamma/2\Delta\sim 0.1$ where
the paramagnetic channel starts to exceed the diamagnetic one.
Contrary to the significant angle dependence for quasiparticles, 
the Higgs-mode contribution quickly becomes isotropic as one deviates from the clean limit.
The angle dependence of the Higgs-mode contribution is no larger than $1.5\%$
at $\gamma/2\Delta\ge 1$.

\begin{figure*}[t]
\includegraphics[width=15cm]{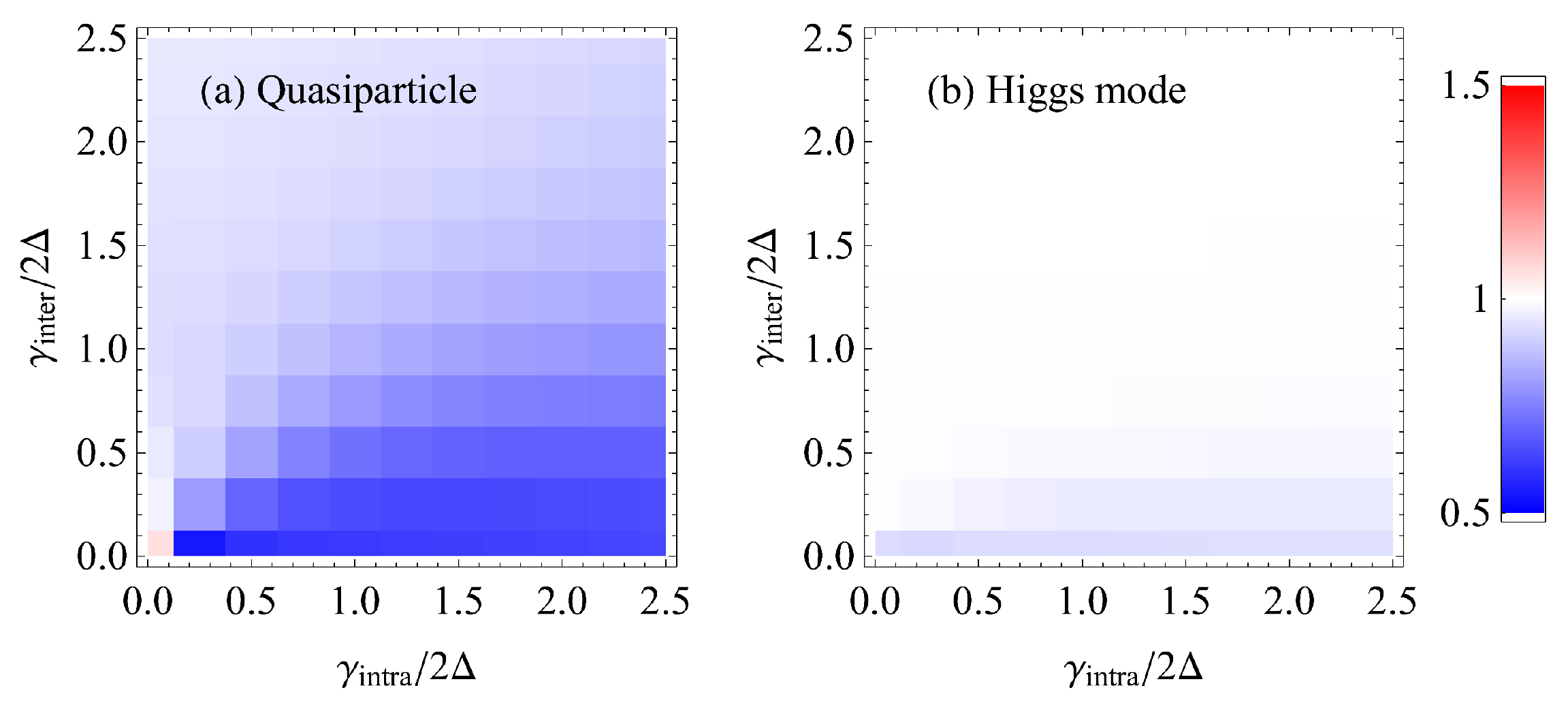}
\caption{The polarization-angle dependence of the quasiparticle (a) and Higgs-mode (b)
contributions $|\chi_{\rm qp, H}(\theta=45^\circ)|^2/|\chi_{\rm qp, H}(\theta=0^\circ)|^2$
to the THG intensity for the model of NbN superconductors (\ref{BCS Hamiltonian}) at frequency $2\Omega=2\Delta$
plotted in the space of the intraband and interband impurity scattering rates.}
\label{fig: THG angle gamma}
\end{figure*}

In Fig.~\ref{fig: THG angle gamma}, we plot $|\chi_{\rm qp}(\theta=45^\circ)|^2/|\chi_{\rm qp}(\theta=0^\circ)|^2$
[Fig.~\ref{fig: THG angle gamma}(a)] and $|\chi_{\rm H}(\theta=45^\circ)|^2/|\chi_{\rm H}(\theta=0^\circ)|^2$ [Fig.~\ref{fig: THG angle gamma}(b)]
in the space of $\gamma_{\rm intra}/2\Delta$ and $\gamma_{\rm inter}/2\Delta$.
The quasiparticle contribution generally shows clear angle dependence for arbitrary impurity scattering rates.
The increasing behavior of $|\chi_{\rm qp}|^2$ as a function of $\theta$ is seen only in the vicinity of the clean limit,
apart from which $|\chi_{\rm qp}|^2$ decreases by $5-50\%$.
It seems that the angle dependence of the quasiparticle contribution does not vanish in the large intraband and/or interband impurity scattering limit.
On the other hand, the angle dependence of the Higgs mode is suppressed 
for general impurity scattering rates as compared to quasiparticles. 
At the vanishing of the interband impurity scattering,
the Higgs-mode contribution shows slight angle dependence of few $\%$.
One can also see that the angle dependence of both the quasiparticle and Higgs-mode contributions is sensitive to the interband impurity scattering rather than the intraband one in the dirty regime.

We expect that there are generally nonvanishing interband impurity scatterings 
in NbN. Then,
one can use the polarization-angle dependence of THG to discriminate the Higgs-mode and quasiparticle contributions in the dirty regime of multiband superconductors.
The optical conductivity measurement \cite{Matsunaga2014, Matsunaga2017} 
suggests that the NbN samples
used in the THG experiment is close to the dirty limit ($\gamma/2\Delta\gg 1$). 
The experimental observation of no angle dependence of THG in NbN superconductors \cite{Matsunaga2017} 
together with our results on the channel-resolved THG intensity in the dirty regime
imply that the dominant contribution to the THG resonance originates from the Higgs mode.

Finally, let us comment on the behavior of the polarization-angle dependence on the ratio $V_{\rm inter}/V_{\rm intra}$. While we used the realistic value of $V_{\rm inter}/V_{\rm intra}=0.18$ for NbN estimated from first principles calculations throughout the paper, we have checked the angle dependence for several other values of $V_{\rm inter}/V_{\rm intra}$ (not shown).
In general, the angle dependence of the Higgs mode tends to be strongly suppressed as one increases $V_{\rm inter}/V_{\rm intra}$ (for the case of $V_{\rm inter}/V_{\rm intra}=1$ in the clean limit, see \cite{Matsunaga2017}), whereas the angle dependence of quasiparticles remains almost unchanged.
Although the realistic value of $V_{\rm inter}/V_{\rm intra}=0.18$ that we obtained in the present paper is not so large, we find that the Higgs-mode contribution is almost polarization-angle independent in the dirty regime for NbN superconductors.
This suggests that the effect of impurity scattering plays an important role in understanding the behavior of the THG resonance in NbN superconductors.

\section{Summary and discussions}
\label{summary}

To summarize, we study the resonance of third harmonic generation and its polarization-angle dependence
in disordered NbN superconductors based on the effective three-band model constructed from
first principles calculations on the electron and phonon band structures of NbN. 
Using the density functional perturbation theory, we evaluate the band-resolved matrix elements
of the electron-phonon coupling constants for NbN, 
and the ratio between the intraband and interband pairing interactions, $V_{\rm inter}/V_{\rm intra}$, is found to be about 0.17-0.18. 

We input the evaluated ratio between the pairing interaction parameters in the effective model,
whose THG susceptibility is calculated in the channel-resolved manner
with the BCS mean-field and self-consistent Born approximations.
The results show that in the dirty regime the dominant contribution to the THG resonance is given by 
the Higgs mode in the paramagnetic channel, which does not have polarization-angle dependence
with nonvanishing interband impurity scattering.
The second dominant one is given by quasiparticles in the paramagnetic channel, which exhibit
clear polarization-angle dependence in the dirty regime. Our results are quite consistent
with the polarization-resolved THG experiment on NbN superconductors \cite{Matsunaga2017}, 
which have found no polarization-angle dependence in the THG resonance.
It will be interesting if one can test the impurity dependence of the THG by controlling impurity concentration in NbN in future experiments.

While we have focused on Higgs amplitude mode in the present paper, there could arise the collective phase mode coupled to electromagnetic fields at low energies in disordered superconductors \cite{CarlsonGoldman1973,CarlsonGoldman1975,Kulik1981}.
Since (i) the mode energy is generally different from $2\Delta$, (ii) it can exist only in the vicinity of $T_c$, and (iii)
the phase mode is decoupled from the amplitude mode when an approximate particle-hole symmetry is present (as is the case in the BCS approximation), we expect that the phase mode (if it may exist) will not affect the THG resonance observed at frequency being half of $2\Delta$.
In fact, such a phase mode has not been observed as the THG resonance in experiments. However, it would be worthwhile to pursue a possibility of detecting the low-energy phase mode by nonlinear optical responses in the future.

Our scheme of classifying and calculating THG susceptibilities for disordered superconductors 
from first principles can be applied
not only to NbN but also to other superconductors. Interesting future applications include
the THG resonance in MgB$_2$, a multigap superconductor with multiple Higgs modes as well as
the Leggett mode \cite{Leggett1966, Akbari2013, Krull2016, Murotani2017}, and NbSe$_2$, where superconductivity and charge density wave coexist.
For unconventional superconductors such as cuprates \cite{BarlasVarma2013, Katsumi2018, Schwarz2020, Chu2020, Katsumi2019, SchwarzManske2020} and iron-based superconductors, 
we need to extend the present formalism to take into account strong correlation effects
beyond the BCS approximation in THG,
which we leave as a future problem.

\acknowledgements

We acknowledge R. Shimano and Y. Murotani for valuable discussions.
We thank various discussions at the international conference of Ultrafast and
Nonlinear Dynamics of Quantum Materials (Paris Ultrafast 2019),
where part of the present work has been presented.
N.T. acknowledges support by JSPS KAKENHI (Grants No. JP16K17729, No. JP20K03811)
and JST PRESTO (Grant No. JPMJPR16N7).
Y.N. is supported by JSPS KAKENHI (Grants No. JP16H06345, No. JP17K14336 and No. JP18H01158).

\appendix

\begin{widetext}

\section{Derivation of Eq.~(\ref{attraction})}
\label{appendix: FS average}

In this appendix, we show the derivation of the Fermi-surface average of the effective phonon-mediated attractive interaction [Eq.~(\ref{attraction}) in the main text].
As we discussed in Sec.~\ref{el-ph coupling}, the momentum dependent effective interaction is given by 
$V_{mn}(\bm k,\bm q)=\sum_\nu |g_{mn}^\nu(\bm k,\bm q)|^2 \frac{2}{\omega_{\bm q\nu}}$.
Here $m$ and $n$ ($= xy, yz, zx$) represent the indices for maximally localized Wannier orbitals that we construct from the first principles band structure calculations.
In the case of NbN, off-diagonal hopping matrix elements are negligibly small between different Wannier orbitals. If we neglect the off-diagonal components and if we appropriately choose the ordering of the orbital indices, the Fermi surface average of the effective interaction $V_{mn}(\bm k,\bm q)$ can be defined by
\begin{align}
\langle V_{mn} \rangle_{\rm FS}
&=
\frac{\sum_{\bm k, \bm q}
\delta(\epsilon_{\bm k+\bm q m}-\epsilon_F)
\delta(\epsilon_{\bm k n}-\epsilon_F)  
V_{mn}(\bm k,\bm q)} 
{\sum_{\bm k,\bm q}
\delta(\epsilon_{\bm k+\bm q m}-\epsilon_F)
\delta(\epsilon_{\bm k n}-\epsilon_F)}
=
\frac{1}{N_{\bm k}N_{\bm q}}
\sum_{\bm k, \bm q}
\frac{\delta(\epsilon_{\bm k+\bm q m}-\epsilon_F)
\delta(\epsilon_{\bm k n}-\epsilon_F)} 
{D^2(\epsilon_F)}V_{mn}(\bm k,\bm q),
\label{Eq_FS_av_Bloch}
\end{align}
where the density of states at the Fermi energy for orbital $n$ is given by
\begin{align}
D(\epsilon_F)
&=
\frac{1}{N_{\bm k}}
\sum_{\bm k} \delta(\epsilon_{\bm k n}-\epsilon_F).
\end{align}
Note that the density of states $D(\epsilon_F)$ does not depend on $n$ due to the symmetry among the $t_{2g}$
orbitals for NbN.

When off-diagonal hoppings in the Wannier basis are not negligible,
Eq.~(\ref{Eq_FS_av_Bloch}) is not directly applicable 
since the orbitals are highly mixed in the band basis near degenerate ${\bm k}$ points.
To see this, let us explicitly write the Hamiltonian in the Wannier basis,
\begin{align}
H
&=
\sum_{\bm kmn\sigma} c_{\bm km\sigma}^\dagger H_{mn}^{\rm Wannier}(\bm k) c_{\bm kn\sigma},
\end{align}
with the diagonal elements $H_{nn}^{\rm Wannier}(\bm k)=\epsilon_{\bm kn}$.
If one goes to the band basis, the Hamiltonian becomes diagonal, 
\begin{align}
H
&=
\sum_{\bm k\alpha\beta\sigma} c_{\bm k\alpha\sigma}^\dagger H_{\alpha\beta}^{\rm band}(\bm k) c_{\bm k\beta\sigma},
\quad
H_{\alpha\beta}^{\rm band}(\bm k)
=
\epsilon_{\bm k\alpha}\delta_{\alpha\beta},
\end{align}
where $\alpha$ and $\beta$ are the Bloch band indices. $H^{\rm Wannier}$ and $H^{\rm band}$ are related through a unitary transformation,
\begin{align}
H_{mn}^{\rm Wannier}(\bm k)
&=
\sum_{\alpha\beta} U_{m\alpha}^{\bm k}H_{\alpha\beta}^{\rm band}(\bm k)(U_{n\beta}^{\bm k})^\ast.
\end{align}
When there are off-diagonal hoppings between different orbitals, 
the level repulsion occurs and $\epsilon_{\bm k\alpha}$ does not coincide 
with $\epsilon_{\bm kn}$ in general.
To correctly describe the spectral weight in multi-orbital systems, we use the retarded Green’s function, 
\begin{align}
G_{\bm kmn}^R(\omega)
&=
(\omega+i\epsilon-H^{\rm Wannier}(\bm k))^{-1}_{mn}
\notag
\\
&=
(\omega+i\epsilon-U^{\bm k} H^{\rm band}(\bm k)U^{\bm k\dagger})^{-1}_{mn}
\notag
\\
&=
\sum_\alpha U_{m\alpha}^{\bm k} \frac{1}{\omega+i\epsilon-\epsilon_{\bm k\alpha}}
(U_{n\alpha}^{\bm k})^\ast.
\end{align}
The spectral function in multi-orbital systems is given by the imaginary part of the retarded Green’s function,
\begin{align}
A_{nn}(\bm k,\omega)
&=
-\frac{1}{\pi}{\rm Im}\, G_{\bm knn}^R(\omega)
=
\sum_\alpha |U_{n\alpha}^{\bm k}|^2 \delta(\omega-\epsilon_{\bm k\alpha}).
\end{align}
Then, the spectral weight of orbital $n$ at the Fermi energy is given by
\begin{align}
w_{\bm kn}
&=
A_{nn}(\bm k,\epsilon_F)
=
\sum_\alpha |U_{n\alpha}^{\bm k}|^2 \delta(\epsilon_{\bm k\alpha}-\epsilon_F).
\label{w_kn general}
\end{align}
If the off-diagonal components in the Wannier basis are absent, and
if one chooses the ordering of band indices to match the orbital indices,
the unitary matrix becomes identity, $U_{n\alpha}^{\bm k}=\delta_{n\alpha}$,
and the weight becomes $w_{\bm kn}=\delta(\epsilon_{\bm kn}-\epsilon_F)$.
The density of states at the Fermi energy for orbital $n$ is given by
\begin{align}
D(\epsilon_F)
&=
\frac{1}{N_{\bm k}} \sum_{\bm k} w_{\bm kn}
=
\frac{1}{N_{\bm k}} \sum_{\bm k} A_{nn}(\bm k,\epsilon_F)
=
\frac{1}{N_{\bm k}} \sum_{\bm k\alpha} |U_{n\alpha}^{\bm k}|^2 \delta(\epsilon_{\bm k\alpha}-\epsilon_F).
\label{DOS}
\end{align}
In the case of NbN, the density of states does not depend on $n$ due to the reason stated above.


Using the general expression of the spectral weight $w_{\bm kn}$ (\ref{w_kn general}),
the Fermi surface average of the effective interaction is defined by
\begin{align}
\langle V_{mn} \rangle_{\rm FS}
&=
\frac{\sum_{\bm k, \bm q} 
w_{\bm k+\bm q m}w_{\bm kn}
V_{mn}(\bm k, \bm q)}
{\sum_{\bm k, \bm q} w_{\bm k+\bm q m}w_{\bm k n}}.
\end{align}
This definition can be obtained by simply replacing the delta functions in Eq.~(\ref{Eq_FS_av_Bloch}) with $w_{\bm kn}$.
By using the density of states (\ref{DOS}) and assuming that the density of states does not depend on orbitals,
we arrive at
\begin{align}
\langle V_{mn} \rangle_{\rm FS}
&=
\frac{1}{N_{\bm k}N_{\bm q}}\sum_{\bm k, \bm q} 
\frac{w_{\bm k+\bm q m}w_{\bm kn}}{D^2(\epsilon_F)}
V_{mn}(\bm k,\bm q).
\label{general FS average}
\end{align}
This is the general expression for the Fermi-surface-averaged effective interaction
that we showed as Eq.~(\ref{attraction}) in the main text. 
In our calculations, we find that NbN has very small off-diagonal hoppings, where Eqs.~(\ref{general FS average})
and (\ref{Eq_FS_av_Bloch}) are almost equivalent. Even in this case, however, it is safe to use Eq.~(\ref{general FS average})
since the ordering of the orbital indices may be shuffled at various $\bm k$ points (so that it becomes tiresome to track the label ordering) 
and in the vicinity of degenerate points
($\epsilon_{\bm km}\approx \epsilon_{\bm kn}$ for $m\neq n$) off-diagonal components may not be neglected.

\section{THG susceptibilities for disordered multiband superconductors}
\label{appendix: THG}

In this Appendix, we present the detailed formulation of THG susceptibilities for disordered multiband superconductors
within the BCS mean-field and self-consistent Born approximations.
The basic idea of the derivation has been given in Sec.~\ref{method} in the main text.
We differentiate the Green's function, the self-energy, and the superconducting gap with respect to the external field,
and determine the second-order derivatives, $\ddot{\hat G}_{\bm kn}(\omega)$ (\ref{ddot G}), $\ddot{\hat\Sigma}_n(\omega)$ (\ref{ddot self-energy}), and $\ddot\Delta_n$ (\ref{ddot gap}), in the self-consistent manner.

To this end, we first determine the $\tau_1$ vertex function $\hat\Lambda_{mn}^{\tau_1}(\omega;\Omega)$,
which is the $\tau_1$ vertex dressed by the impurity-ladder corrections,
satisfying the Bethe-Salpeter equation,
\begin{align}
\hat\Lambda_{mn}^{\tau_1}(\omega;\Omega)
&=
\hat\tau_1\delta_{mn}+
\sum_l\gamma_{ml}^2 \frac{1}{N_{\bm k}}\sum_{\bm k}
\hat\tau_3 \hat G_{\bm kl}(\omega+2\Omega)\hat\Lambda_{ln}^{\tau_1}(\omega;\Omega)
\hat G_{\bm kl}(\omega)\hat\tau_3,
\label{appendix: BS tau1}
\end{align}
where we omit the superscript $\alpha=R,A,<$. We also have the Bethe-Salpeter equations for
the diamagnetic and paramagnetic vertex functions $\hat\Lambda_n^{\rm dia}(\bm k,\omega;\Omega)$
and $\hat\Lambda_n^{\rm para}(\bm k,\omega;\Omega)$,
\begin{align}
\hat\Lambda_n^{\rm dia}(\bm k,\omega;\Omega)
&=
\ddot\epsilon_{\bm kn}\hat\tau_3+
\sum_m \gamma_{nm}^2 \frac{1}{N_{\bm k}}\sum_{\bm k}
\hat\tau_3 \hat G_{\bm km}(\omega+2\Omega)\hat\Lambda_m^{\rm dia}(\bm k,\omega;\Omega)
\hat G_{\bm km}(\omega)\hat\tau_3,
\label{appendix: BS dia}
\\
\hat\Lambda_n^{\rm para}(\bm k,\omega;\Omega)
&=
2\dot\epsilon_{\bm kn}\hat G_{\bm kn}(\omega+\Omega)\dot\epsilon_{\bm kn}
+\sum_m \gamma_{nm}^2 \frac{1}{N_{\bm k}}\sum_{\bm k}
\hat\tau_3 \hat G_{\bm km}(\omega+2\Omega)\hat\Lambda_m^{\rm para}(\bm k,\omega;\Omega)
\hat G_{\bm km}(\omega)\hat\tau_3.
\label{appendix: BS para}
\end{align}
One can decouple the momentum dependence of the diamagnetic and paramagnetic vertex functions as
\begin{align}
\hat\Lambda_n^{\rm dia}(\bm k,\omega;\Omega)
&=
\ddot\epsilon_{\bm kn}\hat\tau_3+
\hat\Lambda_n^{\rm dia}(\omega;\Omega),
\\
\hat\Lambda_n^{\rm para}(\bm k,\omega;\Omega)
&=
2\dot\epsilon_{\bm kn}\hat G_{\bm kn}(\omega+\Omega)\dot\epsilon_{\bm kn}
+\hat\Lambda_n^{\rm para}(\omega;\Omega).
\end{align}
We solve the self-consistent equations (\ref{appendix: BS tau1}), (\ref{appendix: BS dia}), and (\ref{appendix: BS para})
for $\hat\Lambda_{mn}^{\tau_1}(\omega;\Omega)$, $\hat\Lambda_n^{\rm dia}(\omega;\Omega)$, and $\hat\Lambda_n^{\rm para}(\omega;\Omega)$ numerically by matrix inversion. After that, we substitute them to the following self-consistent equations for the second-order derivatives of the superconducting gap function in the diamagnetic and paramagnetic channels,
\begin{align}
\ddot\Delta_n^{\rm dia}(\Omega)
&=
\frac{i}{2}\sum_m V_{nm}\int \frac{d\omega}{2\pi} \frac{1}{N_{\bm k}}\sum_{\bm k}
{\rm Tr}\left[\hat\tau_1 \hat G_{\bm km}(\omega+2\Omega)\left(\hat\Lambda_m^{\rm dia}(\bm k,\omega;\Omega)
+\sum_l\hat\Lambda_{ml}^{\tau_1}(\omega;\Omega)\ddot\Delta_l^{\rm dia}(\Omega)\right) \hat G_{\bm km}(\omega)\right]^<,
\\
\ddot\Delta_n^{\rm para}(\Omega)
&=
\frac{i}{2}\sum_m V_{nm}\int \frac{d\omega}{2\pi} \frac{1}{N_{\bm k}}\sum_{\bm k}
{\rm Tr}\left[\hat\tau_1 \hat G_{\bm km}(\omega+2\Omega)\left(
\hat\Lambda_m^{\rm para}(\bm k,\omega;\Omega)
+\sum_l\hat\Lambda_{ml}^{\tau_1}(\omega;\Omega)\ddot\Delta_l^{\rm para}(\Omega)\right) \hat G_{\bm km}(\omega)\right]^<,
\end{align}
which are again solved numerically by matrix inversion.

Finally, the THG susceptibility is determined from the vertex functions. The classification of the THG susceptibility in terms of the diagrammatic topology has been given in Table~\ref{THG diagram}. The explicit expressions for the quasiparticle contributions to the THG in each channel are given by 
\begin{align}
\chi_{\rm qp}^{(1)}(\Omega)
&=
\frac{i}{6}\int\frac{d\omega}{2\pi} \frac{1}{N_{\bm k}} \sum_{\bm kn} 
{\rm Tr}[\ddddot\epsilon_{\bm kn}\hat\tau_3 \hat G_{\bm kn}(\omega)]^<,
\\
\chi_{\rm qp}^{(2)}(\Omega)
&=
\frac{i}{2}\int\frac{d\omega}{2\pi} \frac{1}{N_{\bm k}} \sum_{\bm kn} 
{\rm Tr}[\dddot\epsilon_{\bm kn} \hat G_{\bm kn}(\omega+\Omega)\dot\epsilon_{\bm kn} \hat G_{\bm kn}(\omega)]^<
+\frac{i}{6}\int\frac{d\omega}{2\pi} \frac{1}{N_{\bm k}} \sum_{\bm kn} 
{\rm Tr}[\dot\epsilon_{\bm kn} \hat G_{\bm kn}(\omega+3\Omega)\dddot\epsilon_{\bm kn} \hat G_{\bm kn}(\omega)]^<,
\\
\chi_{\rm qp}^{(3)}(\Omega)
&=
\frac{i}{2}\int\frac{d\omega}{2\pi} \frac{1}{N_{\bm k}} \sum_{\bm kn} 
{\rm Tr}[\ddot\epsilon_{\bm kn}\hat\tau_3 \hat G_{\bm kn}(\omega+2\Omega)
\ddot\epsilon_{\bm kn}\hat\tau_3 \hat G_{\bm kn}(\omega)]^<
\notag
\\
&\quad
+\frac{i}{2}\int\frac{d\omega}{2\pi} \frac{1}{N_{\bm k}} \sum_{\bm kn} 
{\rm Tr}[\ddot\epsilon_{\bm kn}\hat\tau_3 \hat G_{\bm kn}(\omega+2\Omega)
\hat\Lambda_n^{\rm dia}(\omega;\Omega) \hat G_{\bm kn}(\omega)]^<,
\\
\chi_{\rm qp}^{(4)}(\Omega)
&=
i\int\frac{d\omega}{2\pi} \frac{1}{N_{\bm k}} \sum_{\bm kn} 
{\rm Tr}[\ddot\epsilon_{\bm kn}\hat\tau_3 \hat G_{\bm kn}(\omega+2\Omega)
\dot\epsilon_{\bm kn} \hat G_{\bm kn}(\omega+\Omega)
\dot\epsilon_{\bm kn} \hat G_{\bm kn}(\omega)]^<
\notag
\\
&\quad
+\frac{i}{2}\int\frac{d\omega}{2\pi} \frac{1}{N_{\bm k}} \sum_{\bm kn} 
{\rm Tr}[\dot\epsilon_{\bm kn} \hat G_{\bm kn}(\omega+3\Omega)
\dot\epsilon_{\bm kn} \hat G_{\bm kn}(\omega+2\Omega) 
\ddot\epsilon_{\bm kn}\hat\tau_3 \hat G_{\bm kn}(\omega)]^<
\notag
\\
&\quad
+\frac{i}{2}\int\frac{d\omega}{2\pi} \frac{1}{N_{\bm k}} \sum_{\bm kn} 
{\rm Tr}[\dot\epsilon_{\bm kn} \hat G_{\bm kn}(\omega+3\Omega)\ddot\epsilon_{\bm kn}\hat\tau_3
\hat G_{\bm kn}(\omega+\Omega) \dot\epsilon_{\bm kn}\hat G_{\bm kn}(\omega)]^<
\notag
\\
&\quad
+i\int\frac{d\omega}{2\pi} \frac{1}{N_{\bm k}} \sum_{\bm kn} 
{\rm Tr}[\ddot\epsilon_{\bm kn}\hat\tau_3 \hat G_{\bm kn}(\omega+2\Omega)
\hat\Lambda_n^{\rm para}(\omega;\Omega) \hat G_{\bm kn}(\omega)]^<
\notag
\\
&\quad
+\frac{i}{2}\int\frac{d\omega}{2\pi} \frac{1}{N_{\bm k}} \sum_{\bm kn} 
{\rm Tr}[\dot\epsilon_{\bm kn} \hat G_{\bm kn}(\omega+3\Omega)
\dot\epsilon_{\bm kn} \hat G_{\bm kn}(\omega+2\Omega) 
\hat\Lambda_n^{\rm dia}(\omega;\Omega) 
\hat G_{\bm kn}(\omega)]^<
\notag
\\
&\quad
+\frac{i}{2}\int\frac{d\omega}{2\pi} \frac{1}{N_{\bm k}} \sum_{\bm kn} 
{\rm Tr}[\dot\epsilon_{\bm kn} \hat G_{\bm kn}(\omega+3\Omega)
\hat\Lambda_n^{\rm dia}(\omega+\Omega;\Omega) 
\hat G_{\bm kn}(\omega+\Omega) \dot\epsilon_{\bm kn}\hat G_{\bm kn}(\omega)]^<,
\\
\chi_{\rm qp}^{(5)}(\Omega)
&=
i\int\frac{d\omega}{2\pi} \frac{1}{N_{\bm k}} \sum_{\bm kn} 
{\rm Tr}[\dot\epsilon_{\bm kn} \hat G_{\bm kn}(\omega+3\Omega)
\dot\epsilon_{\bm kn} \hat G_{\bm kn}(\omega+2\Omega)
\dot\epsilon_{\bm kn} \hat G_{\bm kn}(\omega+\Omega)
\dot\epsilon_{\bm kn} \hat G_{\bm kn}(\omega)]^<
\notag
\\
&\quad
+i\int\frac{d\omega}{2\pi} \frac{1}{N_{\bm k}} \sum_{\bm kn} 
{\rm Tr}[\dot\epsilon_{\bm kn} \hat G_{\bm kn}(\omega+3\Omega)
\dot\epsilon_{\bm kn} \hat G_{\bm kn}(\omega+2\Omega)
\hat\Lambda_n^{\rm para}(\omega;\Omega) \hat G_{\bm kn}(\omega)]^<
\notag
\\
&\quad
+i\int\frac{d\omega}{2\pi} \frac{1}{N_{\bm k}} \sum_{\bm kn} 
{\rm Tr}[\dot\epsilon_{\bm kn} \hat G_{\bm kn}(\omega+3\Omega)
\hat\Lambda_n^{\rm para}(\omega+\Omega;\Omega) \hat G_{\bm kn}(\omega+\Omega)
\dot\epsilon_{\bm kn} \hat G_{\bm kn}(\omega)]^<.
\end{align}
The expressions for the Higgs-mode contributions are given by
\begin{align}
\chi_{\rm H}^{(1)}(\Omega)
&=
\frac{i}{2}\int\frac{d\omega}{2\pi} \frac{1}{N_{\bm k}} \sum_{\bm kn} 
{\rm Tr}[\ddot\epsilon_{\bm kn}\hat\tau_3 \hat G_{\bm kn}(\omega+2\Omega)
\ddot\Delta_n^{\rm dia}(\Omega)\hat\tau_1 \hat G_{\bm kn}(\omega)]^<,
\\
\chi_{\rm H}^{(2)}(\Omega)
&=
\frac{i}{2}\int\frac{d\omega}{2\pi} \frac{1}{N_{\bm k}} \sum_{\bm kn} 
{\rm Tr}[\ddot\epsilon_{\bm kn}\hat\tau_3 \hat G_{\bm kn}(\omega+2\Omega)
\ddot\Delta_n^{\rm para}(\Omega)\hat\tau_1 \hat G_{\bm kn}(\omega)]^<
\notag
\\
&\quad
+\frac{i}{2}\int\frac{d\omega}{2\pi} \frac{1}{N_{\bm k}} \sum_{\bm kn} 
{\rm Tr}[\dot\epsilon_{\bm kn} \hat G_{\bm kn}(\omega+3\Omega)
\dot\epsilon_{\bm kn} \hat G_{\bm kn}(\omega+2\Omega) 
\ddot\Delta_n^{\rm dia}(\Omega) \hat\tau_1
\hat G_{\bm kn}(\omega)]^<
\notag
\\
&\quad
+\frac{i}{2}\int\frac{d\omega}{2\pi} \frac{1}{N_{\bm k}} \sum_{\bm kn} 
{\rm Tr}[\dot\epsilon_{\bm kn} \hat G_{\bm kn}(\omega+3\Omega)
\ddot\Delta_n^{\rm dia}(\Omega) \hat\tau_1
\hat G_{\bm kn}(\omega+\Omega) \dot\epsilon_{\bm kn}\hat G_{\bm kn}(\omega)]^<,
\\
\chi_{\rm H}^{(3)}(\Omega)
&=
\frac{i}{2}\int\frac{d\omega}{2\pi} \frac{1}{N_{\bm k}} \sum_{\bm kn} 
{\rm Tr}[\dot\epsilon_{\bm kn} \hat G_{\bm kn}(\omega+3\Omega)
\dot\epsilon_{\bm kn} \hat G_{\bm kn}(\omega+2\Omega) 
\ddot\Delta_n^{\rm para}(\Omega) \hat\tau_1
\hat G_{\bm kn}(\omega)]^<
\notag
\\
&\quad
+\frac{i}{2}\int\frac{d\omega}{2\pi} \frac{1}{N_{\bm k}} \sum_{\bm kn} 
{\rm Tr}[\dot\epsilon_{\bm kn} \hat G_{\bm kn}(\omega+3\Omega)
\ddot\Delta_n^{\rm para}(\Omega) \hat\tau_1
\hat G_{\bm kn}(\omega+\Omega) \dot\epsilon_{\bm kn}\hat G_{\bm kn}(\omega)]^<.
\end{align}
\end{widetext}

\bibliography{ref}

\end{document}